\documentclass[aps,pre,twocolumn,groupedaddress]{revtex4-2}
\usepackage{amsmath}
\usepackage{graphicx}
\usepackage{natbib}
\usepackage{bm}
\usepackage{color}

\newcommand{\bec}[1]{\mbox{\boldmath $ #1$}}

{}
\newcommand{\meanUU}{\overline{\bm{U}}}

\newcommand{\meanWW}{\overline{\mbox{\boldmath $W$}}{}}{}
{}

\newcommand{\meanU}{\overline{U}}

\newcommand{\meanTheta}{\overline{\Theta}}

\def\half{{\textstyle{1\over2}}}

\begin{document}
\title{Semi-organized structures and turbulence in the atmospheric convection}
\author{I.~Rogachevskii$^{1,2}$}
\email{gary@bgu.ac.il}
\homepage{http://www.bgu.ac.il/~gary}
\author{N. Kleeorin$^{1,3}$}
\email{nat@bgu.ac.il}
\bigskip
\affiliation{
$^1$Department of Mechanical Engineering, Ben-Gurion University of the Negev, Beer-Sheva
 8410530, Israel
 \\
$^2$Nordita, Stockholm University and KTH Royal Institute of Technology, 10691 Stockholm, Sweden
\\
$^{3}$ Institute of Continuous Media Mechanics, Korolyov str. 1, Perm  614013, Russia
}

\date{\today}
\begin{abstract}
The atmospheric convective boundary layer (CBL) consists of three basic parts: (i) the surface layer unstably stratified and dominated by small-scale turbulence of very complex nature; (ii) the CBL core dominated by the energy-, momentum- and mass-transport of semi-organized structures (large-scale circulations), with a small contribution from small-scale turbulence produced by local structural shears; and (iii) turbulent entrainment layer at the upper boundary, characterized by essentially stable stratification with negative (downward) turbulent flux of potential temperature.
The energy- and flux budget (EFB) theory developed previously for atmospheric stably-stratified turbulence
and the surface layer in atmospheric convective turbulence is extended to the CBL core
using budget equations for turbulent energies and turbulent fluxes of buoyancy and momentum.
For the CBL core, we determine global turbulent characteristics
(averaged over the entire volume of the semi-organized structure) as well as
kinetic and thermal energies of the semi-organized structures
as the functions of the aspect ratio of the semi-organized structure, the scale separation parameter between
the vertical size of the structures and the integral scale of turbulence
and the degree of thermal anisotropy characterized the form of plumes.
The obtained theoretical relationships are potentially useful in modeling applications
in the atmospheric convective boundary-layer,
and analysis of laboratory and field experiments,
direct numerical simulations and large-eddy simulations of convective turbulence
with large-scale semi-organized structures.
\end{abstract}

%e-print: NORDITA-2023-073

\maketitle

\section{Introduction}
\label{sec:1}

Conventional theory of turbulence and methods of calculation of turbulent transport coefficients
are based on the following classical paradigm \citep[see, e.g., Refs.][]{MY13,LE08,SC08,D13,RI21}.
Turbulent flow represents a superposition of the two types of motion: fully organized mean flow
and fully chaotic turbulence produced, e.g.,  by the mean-flow velocity shears.
Turbulence comprises an ensemble of chaotic motions (turbulent eddies) of different scales
characterized by the forward energy cascade from larger to smaller eddies.
The spectrum of turbulence has an inertial range characterized by the energy flux
towards smaller eddies with constant energy dissipation rate.
The energy flux is balanced by the viscous dissipation at the smallest eddies at the viscous range of scales.

Accordingly, the mean flow is treated deterministically, while turbulence is described statistically.
The local characteristics of turbulence (in particular, turbulent fluxes that appear in the Reynolds-averaged equations)
are controlled by local features of the mean flow.
The turbulent flux of any transporting property is proportional to the mean gradient of the property
multiplied by appropriate turbulent-exchange coefficient.
This concept of down-gradient turbulent transport reduces the turbulence-closure problem
to determining the above exchange coefficients: eddy viscosity $K_{\rm M}$, eddy diffusivity $K_{\rm D}$
and turbulent heat conductivity $K_{\rm H}$ that, in turn, are assumed proportional to the product
of turbulent kinetic energy $E_{\rm K}$ and turbulent timescale $t_{\rm T}$.
The above turbulence paradigm and the concept of down-gradient transport
have been formulated for the shear-generated turbulence in neutrally stratified flows \citep[see, e.g., Refs.][]{MY13,LE08,SC08,D13},
and have proven applicable to a wide range of neutrally and weakly stably or unstably stratified
flows.

However, there is increasing evidence of their poor applicability to both strongly stable stratification
and strongly unstable stratification \citep[see, e.g., Refs.][]{OH01,SF01,RK04,MV05,SR10,ML10,ZKR13,ML14,LK16,L19}.
The present paper is devoted to atmospheric convective boundary layer (CBL), which involves besides
the mean flow and Kolmogorov’s turbulence, the two additional types of motion disregarded in the conventional
theory:
\begin{itemize}
\item{
Small-scale buoyancy-driven vertical plumes, which exhibit inverse energy transfer, namely,
merge to form larger and larger plumes instead of breaking down and feeding kinetic energy
of horizontal velocity fluctuations, as it should be in the case of the forward cascade
\cite{ZRK21};
}
\item{
Large-scale semi-organized convective structures (energetically supplied by merging plums),
which embrace the entire CBL, and perform non-local transports irrespective
of mean gradients of transporting properties
\citep[see, e.g., Refs.][]{ZHE06,EKRZ02,EKRZ06,HZ13}.
}
\end{itemize}

\begin{figure}
\vspace*{1mm}
\centering
\includegraphics[width=8cm]{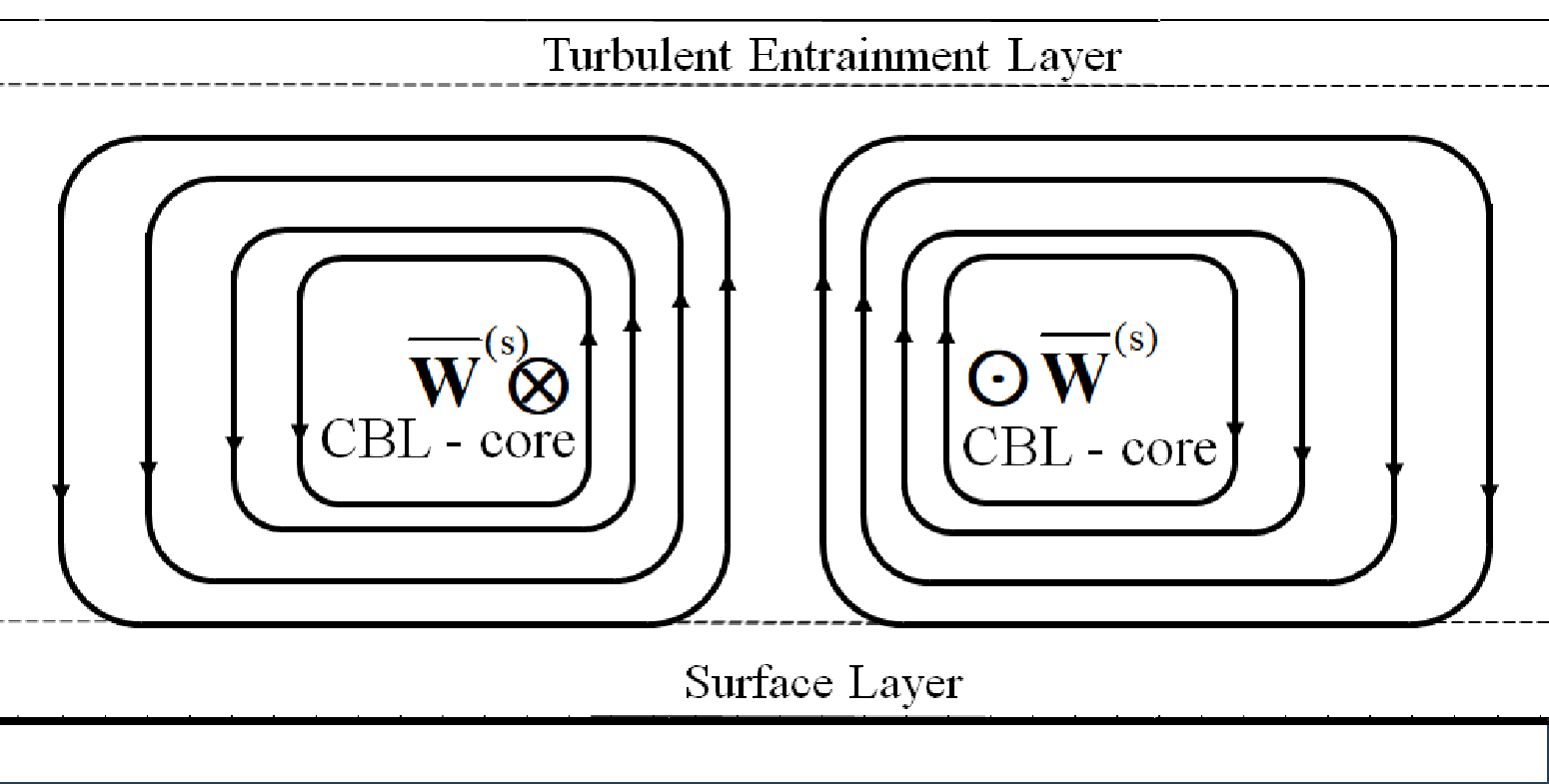}
\caption{\label{Fig1}
Structure of the atmospheric convective boundary layer (CBL),
where $\meanUU^{({\rm s})}$ and $\meanWW^{({\rm s})} =\bec{\nabla} {\bm \times}  \meanUU^{({\rm s})}$ are the
velocity and vorticity characterising the semi-organized structures in CBL.
}
\end{figure}

We recall that the CBL in the atmosphere develops against strongly stable stratification in the free flow.
This leads to the formation of comparatively thin, stably stratified turbulent entrainment layer at the CBL upper boundary.
The turbulent entrainment layer separates CBL from the free atmospheric flow and acts similarly to the upper lid in laboratory
experiments, causing the development of semi-organized structures: large-scale convective cells (the cloud cells) in the shear-free CBL
(analogous to large-scale circulations in laboratory experiments)
and large-scale convective rolls (the cloud streets) in the sheared CBL \citep[see, e.g., Refs. ][]{EKRZ06,HZ13}.
The convective semi-organized structures disturb the CBL-free flow interface, which leads to
exciting internal gravity waves in the free atmospheric flow
and pumping the energy out of CBL \cite{ZKR09,KRZ19}.
Taking into account the above processes, it is convenient to divide CBL into three basic parts
(see Fig.~\ref{Fig1}):
\begin{itemize}
\item{
Shallow surface layer strongly unstably stratified and dominated by vertical transport due to the small-scale three-dimensional turbulence produced by both mean-wind shears and structural convective-wind shears
of semi-organized structures, and strongly anisotropic buoyancy-driven merging-plum turbulence;
}
\item{
Deep CBL core with preferable vertical transport due to the semi-organized structures and small contribution from three-dimensional turbulence produced by local shears of the semi-organized structures, and very small vertical gradient of the mean potential temperature;
}
\item{
Shallow turbulent entrainment layer at the CBL upper boundary with strong stable stratification dominated by turbulent transport and the downward turbulent flux of potential temperature.
}
\end{itemize}

Observations in the atmospheric CBL, laboratory experiments and large-eddy simulation (LES) confirm that convective structures principally differ from turbulent eddies. The characteristic scales of the semi-organized structures are much larger than the integral turbulence scale and their life-times are much longer than the largest turbulent time scales \citep[see, e.g.,][]{EKRZ06,HZ13}.
In the shear-free atmospheric CBL, the semi-organized structures (the cloud cells) are similar to Bernard cells.
They consist of narrow uprising flows surrounded by wide downdraughts, embrace the entire CBL (up to 1-3 km height), and include pronounced convergence flows toward the cell axes in the near-surface layer, as well as divergence flows at the CBL upper boundary.
In the sheared CBL, the semi-organized structures (the cloud streets) have the form of rolls stretched along the mean wind.
Various features in the atmospheric convective turbulence
have been studied theoretically and numerically \citep[see, e.g., Refs.][]{EGKR06,CA08,GD13,BK15,SCB17,SA18,SA20,JB21,LQB22},
and in the field experiments \citep[see, e.g.,][]{ZHE06,SLM16,GP18},
see reviews \cite{W10,T17}, and references therein.

Deterministic treatment of semi-organized convective structures as distinct from turbulence treated statistically has been employed
to derive non-local convective heat/mass-transfer law for the shear-free CBL \cite{ZHE06}.
Interesting features of convective turbulence now attributed to merging-plume mechanism implying inverse energy transfer from smaller to larger plumes were found long ago \cite{ZRK21}.
The ideas that the shear-produced turbulence interacts with convective semi-organized structures in the same way as with usual mean flow
have been used in a number of studies \cite{EKRZ02,EKRZ06}.

In the present paper we focus on physical processes in the CBL core and
extend the energy- and flux budget (EFB) turbulence closure theory
developed previously for atmospheric stably-stratified turbulence \cite{ZKR07,ZKR08,ZKR09,ZKR10,ZKR13,KRZ19},
turbulent transport of passive scalar \cite{KRZ21}
and the surface layers in atmospheric convective turbulence \cite{RKZ22}, to
convective turbulence in the CBL core in shear-free convection with very weak mean wind.

The EFB theory for the stably stratified turbulence
explains the existence of strong turbulence produced
by large-scale shear for any stratification \cite{ZKR07,ZKR08,ZKR09,ZKR10,ZKR13,KRZ19}.
The physics related to self-maintaining of a stably stratified turbulence for any stratification
is caused by the following.
The increase in the buoyancy due to an enhancement of the vertical gradient of the mean potential temperature
results in a conversion of turbulent kinetic energy (TKE) into turbulent potential energy (TPE).
This decreases the negative down-gradient vertical turbulent flux of potential temperature
by a positive non-gradient turbulent flux of potential temperature originated from enhanced TPE.
This mechanism of the self-control feedback decreases the buoyancy
and maintains stably stratified turbulence for any stratification \cite{KRZ21,RKZ22,RI21} in agreement
with wide experimental evidence \cite{OH01,SF01,RK04,SR10,ML10,ML14,L19,CCH20}.

The EFB theory for the surface layers in atmospheric convective turbulence \cite{RKZ22}
describes a smooth transition between a stably-stratified turbulence
and a convective turbulence, providing analytical expressions for the vertical profiles for all turbulent characteristics
in the entire surface layer including TKE, the intensity of  turbulent potential temperature fluctuations,
the vertical turbulent fluxes of momentum and buoyancy (proportional to potential temperature), the integral turbulence scale,
the turbulence anisotropy, the turbulent Prandtl number and the flux Richardson number.
The obtained analytical vertical profiles describe also
the transition range between the lower and upper parts of the surface layer.

The EFB theory for the CBL core developed in the present study, is based on the threefold decomposition:
mean flow, semi-organized strictures and small-scale turbulence, in combination with analytical description
of convective semi-organized structures in the shear-free CBL.
We find the global turbulent characteristics (averaged over the entire volume of the semi-organized structure)
and kinetic and thermal energies of the semi-organized structures, which depend on the aspect ratio of the semi-organized structure
and scale separation parameter between the vertical size of the structures and the integral scale of turbulence.
The obtained theoretical relationships are potentially useful in modeling applications
in the atmospheric convective boundary-layer.

This paper is organized as follows.
In Sec.~\ref{sec:2} we discuss turbulent flux of  potential temperature and its effect on
the formation of semi-organized structures.
In this section we derive expressions for the velocity and potential temperature of convective semi-organized structures.
In Sec.~\ref{sec:3} we study turbulence in the CBL core,
starting with budget equations for turbulent energies and turbulence
fluxes of potential temperature and momentum  (Sec.~\ref{subsec:3.1}),
and formulate main assumptions for the energy and flux budget
turbulence closure theory for convective turbulence (Sec.~\ref{subsec:3.2}).
In the framework of this theory, we derive expressions for the global characteristics
of convective turbulence and semi-organized structures (Sec.~\ref{subsec:3.3}).
In Sec.~\ref{sec:4} we consider a transition from the convective surface layer to the CBL core,
where we perform a matching between the solutions obtained for the convective surface layer and the CBL core.
Finally, in Sec.~\ref{sec:5} we discuss the obtained results and draw conclusions.

\section{Turbulent flux of potential temperature and semi-organized structures}
\label{sec:2}

The convective boundary layer involves three principally different types
of motion:
\begin{itemize}
\item{
regular plain-parallel mean flow homogeneous in the horizontal plain
(coordinates  $x, y$), but heterogeneous in the vertical (coordinate  $z$);}
\item{
vertically and horizontally heterogeneous long-lived CBL-scale semi-organized convective structures; and}
\item{
small-scale turbulence.}
\end{itemize}
We use capital letters with superscript (m) to denote the mean-flow fields of wind:  $\overline{\bm U}^{({\rm m})} = \left(\overline{U}_x^{({\rm m})}, \overline{U}_y^{({\rm m})}, \overline{U}_z^{({\rm m})}\right)$, pressure $\overline{P}^{({\rm m})}$  and potential temperature $\overline{\Theta}^{({\rm m})}$; capital letters with superscript (s) to denote the semi-organized structure fields,
$\overline{\bm U}^{({\rm s})} = \left(\overline{U}_x^{({\rm s})}, \overline{U}_y^{({\rm s})}, \overline{U}_z^{({\rm s})}\right)$,
and  $\overline{P}^{({\rm s})}$, $\overline{\Theta}^{({\rm s})}$; lower-case letters to denote turbulent fields:  ${\bm u}$,  $p$ and $\theta$;  and just capital letters to denote actual (total) fields, e.g., the total velocity is ${\bm U}=\overline{\bm U}^{({\rm m})} + \overline{\bm U}^{({\rm s})} + {\bm u}$,  the total pressure is $P=\overline{P}^{({\rm m})} + \overline{P}^{({\rm s})} + p$ and the total potential temperature is $\Theta=\overline{\Theta}^{({\rm m})} + \overline{\Theta}^{({\rm s})} + \theta$.
In the present study, we consider a shear-free convection with negligible mean flow field, $\overline{\bm U}^{({\rm m})}=0$, but with a non-zero vertical gradient of the mean potential temperature $\nabla_z \overline{\Theta}^{({\rm m})} \not = 0$.

Turbulent fluxes of potential temperature and momentum are defined as  ${\bm F} = \langle {\bm u} \, \theta \rangle$  and  $\tau_{ij} =\langle u_i \, u_j \rangle$, respectively, where angle brackets denote ensemble averaging.
For the sake of definiteness, we restrict our consideration to dry atmosphere,
so that buoyancy is proportional to the potential temperature  $\Theta = T (P_\ast / P)^{1-\gamma^{-1}}$, where
$T$ is the fluid temperature with the reference value $T_\ast$,
$P$ is the fluid pressure with the reference value $P_\ast$
and $\gamma = c_{\rm p}/c_{\rm v}$ is the specific heat ratio.
The familiar down-gradient approximation of the turbulent fluxes of potential temperature and momentum reads:
${\bm F} = - K_{\rm H} {\bm \nabla} \overline{\Theta}$
and  $\tau_{ij} = - K_{\rm M} \, (\nabla_i \overline{U}_j + \nabla_j \overline{U}_i)$,
where  $\overline{\Theta} = \overline{\Theta}^{({\rm m})} + \overline{\Theta}^{({\rm s})}$,
$\overline{\bm U}=\overline{\bm U}^{({\rm s})}$, $K_{\rm H}$ and  $K_{\rm M}$
are turbulent heat conductivity and turbulent viscosity traditionally treated
as scalars \cite{MY13}.

On the other hand, there are long-lived CBL-scale semi-organized convective structures
and the velocity field inside large-scale convective structures
is strongly nonuniform. These nonuniform motions can produce
anisotropic velocity fluctuations which can contribute
to the turbulent flux of potential temperature.
In particular, the classical turbulent flux of potential temperature, ${\bm F} = - K_{\rm H} {\bm \nabla} \overline{\Theta}$, does not
take into account the contribution from anisotropic velocity fluctuations.

\subsection{Turbulent flux of potential temperature}

The contribution to the turbulent flux of potential temperature
from anisotropic velocity fluctuations plays a crucial  role in
the formation of large-scale semi-organized structures in turbulent convection.
Indeed, the turbulent flux of potential temperature
${\bm F}$ which takes into account anisotropic velocity fluctuations reads \cite{EKRZ02,EKRZ06}:
\begin{eqnarray}
{\bm F} = {\bm F}^{\ast} -  {8 \, t_{\rm T} \over 9} \, \left[\alpha \, {\bm
F}_{z}^{\ast} \, {\rm div} \, \meanUU_{\rm h}^{({\rm s})} - {2 \alpha + 3 \over 10} \,
\meanWW^{({\rm s})} {\bm \times} {\bm F}_z^{\ast}\right] ,
\nonumber\\
 \label{BB1}
\end{eqnarray}
where ${\bm F}^{\ast} = - K_{\rm H} \, \bec{\nabla} \meanTheta^{({\rm m})}$
is the classical  background turbulent flux of potential temperature
in the absence of nonuniform large-scale flows, $\alpha$ is the degree of thermal anisotropy
that characterizes the form of plumes and is defined by Eq.~(\ref{TD6}),
$t_{\rm T}$ is the characteristic turbulent time
at the integral turbulent scale, $\meanWW^{({\rm s})} =
\bec{\nabla} {\bm \times}  \meanUU^{({\rm s})}$ is the mean vorticity characterized the semi-organized structure, the mean velocity
$\meanUU^{({\rm s})}=  \meanUU_{\rm h}^{({\rm s})} +  \meanUU_{z}^{({\rm s})}$ that is decomposed into the horizontal
$\meanUU_{\rm h}^{({\rm s})}$ and vertical $\meanUU_{z}^{({\rm s})}$ components.
The new contributions to the turbulent flux of potential temperature are caused by
anisotropic velocity fluctuations and depend on the mean velocity gradients of the nonuniform large-scale flow.
It has been demonstrated in Refs.~\cite{EKRZ02,EKRZ06,EGKR06,OKR22} that
these new contributions cause the excitation of large-scale convective-wind instability and the formation of
large-scale semi-organized structures.

\begin{figure}
\vspace*{1mm}
\centering
\includegraphics[width=8cm]{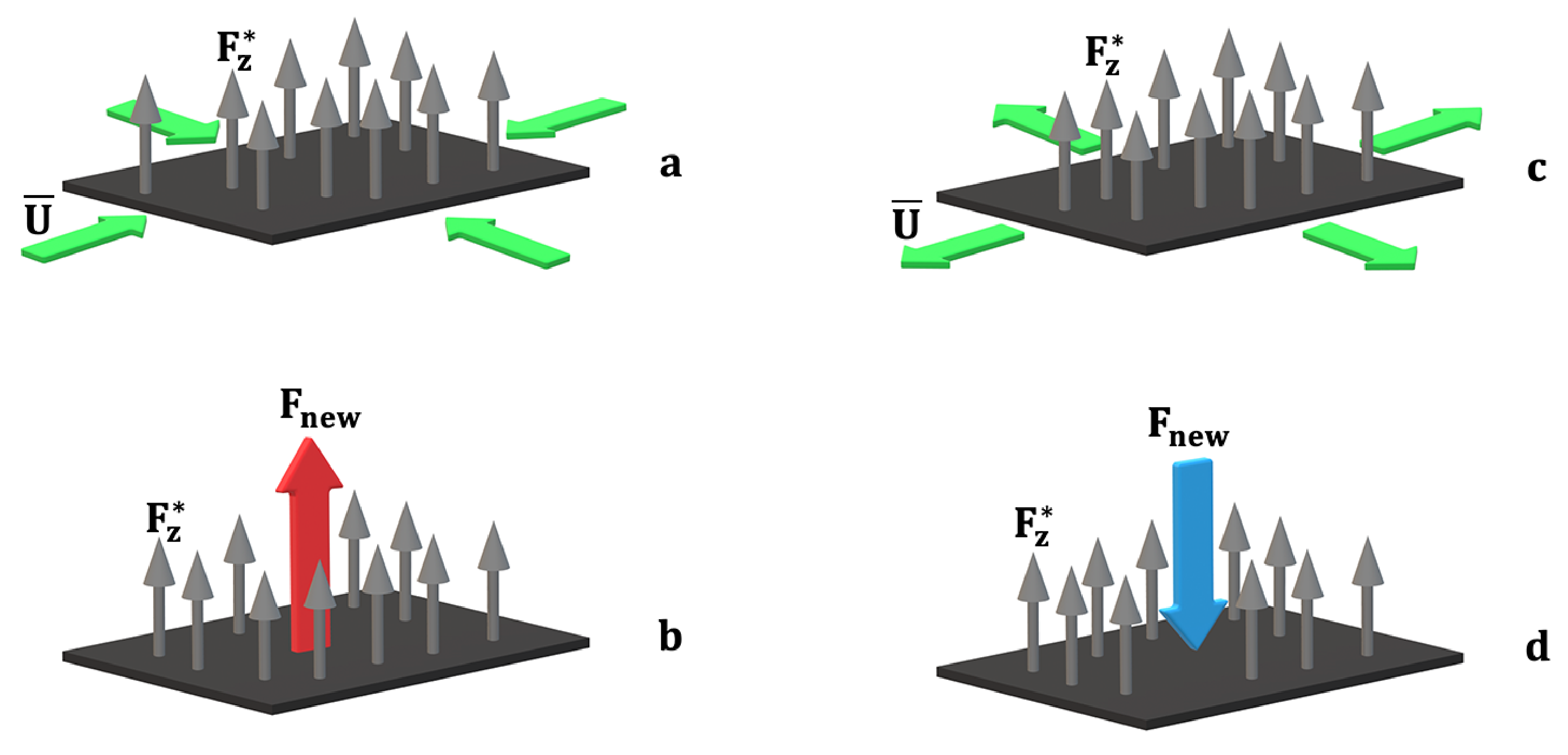}
\caption{\label{Fig2} The mechanism of the large-scale convective-wind instability associated
with the new contribution of the turbulent flux of potential temperature
${\bm F}_{\rm new} = - t_{\rm T} \, \alpha\, {\bm F}_{z}^{\ast} \, {\rm div}
\, \meanUU_{\rm h}^{({\rm s})}$, which
increases (or decreases) the vertical turbulent flux of potential temperature
shown by the red arrow in {\bf b} (or by the blue arrow in {\bf d})
via redistribution of the uniform vertical turbulent flux ${\bm F}_z^{\ast}$
by convergent (or divergent) horizontal mean flows $\meanUU_{\rm h}^{({\rm s})}$
(shown by the green arrows in {\bf a} and {\bf c}).
The vertical turbulent flux ${\bm F}_{\rm new}$ enhances the upward (positive) turbulent flux of potential temperature,
increasing the local mean potential temperature and producing the upward large-scale flow.
Likewise, the vertical turbulent flux ${\bm F}_{\rm new}$ decreases the vertical turbulent flux of potential temperature by the divergent horizontal motions, decreasing the local mean potential temperature and producing the downward large-scale flow.
}
\end{figure}

The mechanism of the large-scale convective-wind instability is related to the second
term ${\bm F}_{\rm new} = - t_{\rm T} \, \alpha\, {\bm F}_{z}^{\ast} \, {\rm div}
\, \meanUU_{\rm h}^{({\rm s})}$ in Eq.~(\ref{BB1}) for the turbulent flux
of potential temperature, which causes the redistribution of the vertical background turbulent
flux of potential temperature ${\bm F}_{z}^{\ast}$ by the perturbations of the
convergent (or divergent) horizontal large-scale flows $\meanUU_{\rm h}^{({\rm s})}$ (see Fig.~\ref{Fig2})
during the life-time of turbulent eddies.
Therefore, this effect increases the vertical turbulent flux of potential temperature by the converging horizontal motions,
which enhances both, the upward (positive) turbulent flux of potential temperature and buoyancy.
The latter forms the upward flow and strengthens the horizontal convergent flow,
resulting in the large-scale convective-wind instability.

On the other hand, the last term $\propto [(2 \alpha + 3) /10] \, t_{\rm T} \, (\meanWW^{({\rm s})} {\bm \times}
{\bm F}_z^{\ast})$ in Eq.~(\ref{BB1}) produces
the horizontal turbulent flux of potential temperature by a
"rotation" of the vertical background turbulent flux ${\bm
F}_{z}^{\ast}$  with the perturbations of the horizontal mean
vorticity $\meanWW_{\rm h}^{({\rm s})}$.
In other words, the contribution to the turbulent flux  of potential temperature
$\propto [(2 \alpha + 3) /10] \, t_{\rm T} \,  (\meanWW^{({\rm s})} {\bm \times}
{\bm F}_z^{\ast})$ creates the horizontal turbulent flux of potential temperature via rotation of
the vertical turbulent flux ${\bm F}_z^{\ast}$ by the  large-scale horizontal vorticity, $\meanWW_{\rm h}^{({\rm s})}$,
decreasing the local potential temperature in rising motions. The latter weakens the buoyancy acceleration,
and  reduces perturbations of the vertical large-scale velocity and vorticity,
contributing to the damping of large-scale convective-wind instability
\cite{EKRZ02,EKRZ06,EGKR06}.

\subsection{Analytical solution for semi-organized structures}

Let us determine the large-scale velocity $\overline{\bm U}^{({\rm s})}$
and potential temperature $\overline{\Theta}^{({\rm s})}$ of the semi-organized structures
formed in small-scale convective turbulence.
Note that the timescale of the growth of the CBL height is much larger than the characteristic time scale of evolution
of the semi-organized structures.
During the formation of these coherent structures there is a two-way nonlinear coupling:
\begin{itemize}
\item{
the effect of small-scale turbulent convection on the formed semi-organized coherent structures; and }
\item{
a back-reaction of the formed semi-organized coherent structures on small-scale turbulent convection.}
\end{itemize}
The velocity and potential temperature inside the semi-organized
structures are strongly non-uniform, causing an anisotropy of convective turbulence. In particular, the convective plumes are extended in vertical direction in regions with strong buoyancy.
To describe such complicated process, we have to solve nonlinear equations for turbulence and mean-field equations for semi-organized structures simultaneously. This cannot be done analytically for large Reynolds and Rayleigh numbers.
To solve this problem, we apply another approach. We use a linearized mean-field equations for the velocity and potential temperature of the semi-organized coherent structures with the parameterized turbulent flux of potential temperature determined by Eq.~(\ref{BB1}). This allows us to describe direct coupling of small-scale convective turbulence with the formed semi-organized coherent structures. However, to take into account back-reaction of the formed semi-organized coherent structures on small-scale convective turbulence, we introduce a phenomenological parameter $\alpha$  that determines a thermal anisotropy and describes an anisotropic form of plumes [see
Eq.~(\ref{TD6}) in Sec.~\ref{subsec:3.3}].
In such phenomenological approach we take into account the back-reaction of the coherent structures on small-scale convective turbulence.

The equations for evolution of the vorticity $\overline{\bm W}^{({\rm s})}$ and potential temperature $\overline{\Theta}^{({\rm s})}$
of semi-organized structures in a fluid flow in the Boussinesq approximation are given by
\begin{eqnarray}
{D \overline{\bm W}^{({\rm s})} \over D t} &=& K_{\rm M} \Delta \overline{\bm W}^{({\rm s})} - \beta ({\bm e} \times {\bm \nabla}) \overline{\Theta}^{({\rm s})}
\nonumber\\
&&
+ \left(\overline{\bm W}^{({\rm s})} \cdot {\bm \nabla}\right) \overline{\bm U}^{({\rm s})} ,
\label{BB2}
\end{eqnarray}
\begin{eqnarray}
{D\overline{\Theta}^{({\rm s})} \over D t} = -  \left({\bm \nabla} \cdot {\bm F} \right) ,
\label{BB3}
\end{eqnarray}
where the turbulent flux of potential temperature ${\bm F}$ is given by Eq.~(\ref{BB1}).
Here ${\bm e}$ is the vertical unit vector, $\beta=g/T_\ast$ is the buoyancy parameter, ${\bm g}$  is the gravity acceleration and
$D / D t = \partial / \partial t + \overline{\bm U}^{({\rm s})} \cdot {\bm \nabla}$.
We restrict our consideration to the quasi-stationary regime and search for axisymmetric ($\vartheta$-independent)
solution to equations for the vorticity $\overline{\bm W}^{({\rm s})}$ and potential temperature $\overline{\Theta}^{({\rm s})}$, in cylindrical coordinates $(r, \vartheta, z)$, where the velocity $\overline{\bm U}^{({\rm s})}$ and vorticity $\overline{\bm W}^{({\rm s})}$ are expressed  in terms of the stream function $\Psi$ as
\begin{eqnarray}
\overline{\bm U}^{({\rm s})} = - {\bm e}_r \, {\partial \Psi \over \partial z} + {\bm e}_z {1 \over r} {\partial (r \, \Psi) \over \partial r} ,
\label{BB4}
\end{eqnarray}

\begin{eqnarray}
\overline{\bm W}^{({\rm s})} = - {\bm e}_\vartheta \, \left(\Delta - r^{-2}\right) \Psi .
\label{BB5}
\end{eqnarray}
Taking the eddy viscosity $K_{\rm M}$ and eddy conductivity $K_{\rm H}$ independent of $r$ and $z$, linearized equations~(\ref{BB2}) and~(\ref{BB3}), we obtain the following steady-state solution:
\begin{eqnarray}
\overline{U}_r^{({\rm s})} = - A_\ast \, \overline{U}_{z0} \, J_1\left({\lambda \, r \over R}\right) \, \cos \left({\pi \, z \over L_z}\right),
\label{BB6}
\end{eqnarray}

\begin{eqnarray}
\overline{U}_z^{({\rm s})} = \overline{U}_{z0} \, J_0\left({\lambda \, r \over R}\right) \, \sin \left({\pi \, z \over L_z}\right),
\label{BB7}
\end{eqnarray}

\begin{eqnarray}
\overline{\Theta}^{({\rm s})} =\overline{\Theta}_{0} \, J_0\left({\lambda \, r \over R}\right) \, \sin \left({\pi \, z \over L_z}\right),
\label{BB8}
\end{eqnarray}
where
\begin{eqnarray}
\Psi = \Psi_{0} \, J_1\left({\lambda \, r \over R}\right) \, \sin \left({\pi \, z \over L_z}\right),
\label{BB9}
\end{eqnarray}

\begin{eqnarray}
\overline{\bm W}^{({\rm s})} = {\bm e}_\vartheta \, {\lambda\, \overline{U}_{z0} \over R} \, \left(1 + A_\ast^2\right) \, J_1\left({\lambda \, r \over R}\right) \, \sin \left({\pi \, z \over L_z}\right) ,
\nonumber\\
\label{BB10}
\end{eqnarray}
and $J_m(x)$ is the Bessel function of the first kind possessing the properties:
$2 J'_m(x)=J_{m-1}(x) - J_{m+1}(x)$ and $J'_0(x) = - J_1(x)$.
Here $\lambda=3.83$  is the first root of equation $J_1(x)=0$,
so that  $J_1(\lambda)=0$, and $A_\ast=\pi \, R / (\lambda \,  L_z)$ is aspect ratio
of the semi-organized structures, where $R$ and $L_z$ are the radius and height of the semi-organized structures, respectively; $\overline{U}_{z0}$, $\overline{\Theta}_{0}$  and $\Psi_{0} = \overline{U}_{z0} \, L_z $  are amplitudes of the velocity, temperature and stream-function, respectively. Substituting Eqs.~(\ref{BB6})--(\ref{BB10}) into Eqs.~(\ref{BB2})--(\ref{BB5}), we obtain  relationships for the amplitudes and the aspect ratio of the semi-organized structures as
\begin{eqnarray}
{\overline{U}_{z0} \over \overline{\Theta}_{0}} = {\beta\, L_z^2 \over \pi^2 \, K_{\rm M}}
\, {A_\ast^2 \over \left(1 + A_\ast^2\right)^2 } ,
\label{BB11}
\end{eqnarray}

\begin{eqnarray}
{K_{\rm M}^2 \over \beta \, F_z \,  t_{\rm T} \, R^2 \, {\rm Pr}_{_{T}}} = {\sigma \, \left(A_\ast^2- \mu\right)
\over \lambda^2 \, \left(1 + A_\ast^2\right)^2 } ,
\label{BB12}
\end{eqnarray}
and $\Psi_{0} = \overline{U}_{z0} \, R / \lambda$,
where ${\rm Pr}_{_{T}}=K_{\rm M} / K_{\rm H}$ is  the turbulent Prandtl number,
$\sigma= 4 \, (8 \alpha-3)/45$ and $\mu=(2 \alpha+3)/(8 \alpha-3)$,
and $\alpha$ is the degree of thermal anisotropy defined by Eq.~(\ref{TD6}) in Sec.~\ref{subsec:3.3}.
The above solution mimic comparatively narrow uprising flow surrounded
by wider and weaker downdraught in reasonable agreement with large-eddy simulations
\citep{HZ13}.

\section{Turbulence in CBL core}
\label{sec:3}

To describe turbulence in the CBL core,
we use the budget equations for the density of turbulent
kinetic energy, the intensity of potential temperature fluctuations
and turbulent fluxes of potential temperature and momentum in the Boussinesq approximation (see, e.g., Ref.~\cite{RKZ22}).
We consider a shear-free turbulent convection with negligibly small mean velocity
$\meanUU^{({\rm m})}$ in comparison with the velocity $\meanUU^{({\rm s})}$ of the semi-organized structures.

\subsection{Budget equations}
\label{subsec:3.1}

We start with the basic equations of the energy and flux budget (EFB) closure theory.
The budget equation for the density of turbulent kinetic energy (TKE), $E_{\rm K}=\langle {\bm u}^2 \rangle/2$,
the intensity of potential temperature fluctuations $E_\theta=\langle \theta^2 \rangle/2$,
the turbulent flux $F_i = \langle u_i \, \theta \rangle$ of potential temperature,
and the Reynolds stress $\tau_{ij} =\langle u_i \, u_j \rangle$ are given by
\begin{eqnarray}
{DE_{\rm K} \over Dt} + \nabla_j \, \Phi_j^{({\rm K})} &=& - \tau_{ij} \, \nabla_j \meanU_i^{({\rm s})}
+ \beta \, F_z - \varepsilon_{_{\rm K}} ,
 \label{C1}
\end{eqnarray}
\begin{eqnarray}
{D E_\theta \over Dt} + \nabla_j \, \Phi_j^{(\theta)} &=& - ({\bm F}{\bf \cdot} \bec{\nabla})\overline{\Theta}^{({\rm s})} - F_z \, \nabla_z \overline{\Theta}^{({\rm m})} - \varepsilon_\theta,
\nonumber\\
 \label{C2}
\end{eqnarray}
\begin{eqnarray}
&&{\partial F_i \over \partial t} + \nabla_j \, {\bm \Phi}_{ij}^{({\rm F})}
= - \tau_{iz} \, \nabla_z  \overline{\Theta}^{({\rm m})} - \tau_{ij} \, \nabla_j \overline{\Theta}^{({\rm s})} + 2 \beta \, E_\theta \, \delta_{i3}
\nonumber\\
&& \qquad- {1 \over \rho_0} \, \langle \theta \, \nabla_i p \rangle - ({\bm F}{\bf \cdot} \bec{\nabla}) \meanU_i^{({\rm s})} - \varepsilon_i^{({\rm F})} ,
\label{C3}
\end{eqnarray}
\begin{eqnarray}
&&{D \tau_{ij} \over Dt} + \nabla_k \, \Phi_{ijk}^{(\tau)} = - \tau_{ik} \, \nabla_k \meanU_j^{({\rm s})} - \tau_{jk} \, \nabla_k \meanU_i^{({\rm s})} + Q_{ij}
\nonumber\\
&& \qquad + \beta \, \left(F_i \delta_{j3} + F_j \delta_{i3}\right) - \varepsilon_{ij}^{(\tau)} ,
 \label{C15}
\end{eqnarray}
where $D / Dt = \partial /\partial t + \meanUU^{({\rm s})} {\bf \cdot} \bec{\nabla}$  is the convective derivative and
$\delta_{ij}$ is the Kronecker unit tensor.
The first term, $- \tau_{i j} \, \nabla_j \meanU_i^{({\rm s})}$, on the RHS of Eq.~(\ref{C1}) is the rate
of production of TKE by the gradients of the velocity  $\meanUU^{({\rm s})}$ of the semi-organized structures.
In particular, turbulence in the CBL-core is produced largely by local shears in semi-organized structures.
We will show that the turbulent kinetic energy is very low compared to kinetic energy of motions in semi-organized structures.
These conclusions are fully confirmed by LES and various laboratory experiments (see, e.g.,  Ref.~\cite{HZ13}, and references therein).
The second production term $\beta \, F_z$ in Eq.~(\ref{C1}) describes buoyancy.
The first two  terms, $- ({\bm F}{\bf \cdot} \bec{\nabla})\overline{\Theta}^{({\rm s})}$ and $- F_z \, \nabla_z \overline{\Theta}^{({\rm m})}$, on the RHS of Eq.~(\ref{C2}) are the rates of production of potential temperature fluctuations,
where the first contribution due to semi-organized structures is dominant one.

The first term, $- \tau_{iz} \, \nabla_z \overline{\Theta}^{({\rm m})}$,  on the RHS of Eq.~(\ref{C3})
contributes to the turbulent flux of potential temperature due to the small vertical gradient of the mean potential temperature,
while the second term, $- \tau_{ij} \, \nabla_j \overline{\Theta}^{({\rm s})}$,
contributes to the turbulent flux caused by the semi-organized structures.
Both terms correspond to the classical gradient mechanism of the turbulent heat transfer.
The third term $2 \beta \, E_\theta \, \delta_{i3}$ on the RHS of Eq.~(\ref{C3}) describes a non-gradient
contribution to the turbulent flux of potential temperature.

The term, $\varepsilon_{_{\rm K}} = \nu \, \langle (\nabla_j u_i)^2 \rangle$, in the RHS of Eq.~(\ref{C1})
is the dissipation rate of the density of the turbulent kinetic energy, where $\nu$ is the kinematic viscosity
of fluid.
The term, $\varepsilon_\theta = \chi \, \langle ({\bm \nabla} \theta)^2 \rangle$ in the RHS of Eq.~(\ref{C2})
is the dissipation rate of the intensity of potential temperature fluctuations $E_\theta$, and $\chi$ is the molecular temperature diffusivity.
The term, $\varepsilon_i^{({\rm F})} = (\nu + \chi) \, \langle (\nabla_j u_i)
\, (\nabla_j \theta) \rangle$ on the RHS of Eq.~(\ref{C3}) is the dissipation rate of the turbulent flux of potential temperature.
The term, $\varepsilon_{ij}^{(\tau)}=2 \nu \, \langle (\nabla_k u_i) \, (\nabla_k u_j) \rangle$ on the RHS of Eq.~(\ref{C15}) is the molecular-viscosity dissipation rate, and the tensor $Q_{ij} = \rho_0^{-1} (\langle p \nabla_i u_j\rangle + \langle p \nabla_j u_i\rangle)$ describes correlations of pressure fluctuations and turbulent velocity gradients.

The term $\Phi_j^{({\rm K})} = \rho_0^{-1} \langle u_j \, p\rangle + (\langle u_j
\, {\bf u}^2 \rangle - \nu \, \nabla_j\langle {\bf u}^2 \rangle)/2$ in Eq.~(\ref{C1}) determines the flux of $E_{\rm K}$,
where $\rho_0$ is the fluid density.
The term $\Phi_j^{(\theta)} = \left(\langle u_j \, \theta^2 \rangle - \chi \, \nabla_j \langle  \theta^2 \rangle\right)/2$
in Eq.~(\ref{C2}) describes the flux of $E_\theta$.
The term  ${\bm \Phi}_{ij}^{({\rm F})} = \langle u_i \, u_j
\, \theta\rangle - \nu \, \langle \theta \, (\nabla_j u_i) \rangle - \chi\, \langle u_i \, (\nabla_j \theta) \rangle$
in Eq.~(\ref{C3})
determines the flux of $F_i$.
The term  $\Phi_{ijk}^{(\tau)}=\langle u_i \, u_j \, u_k\rangle + \rho_0^{-1} \, (\langle p \, u_i\rangle \delta_{jk} + \langle p \, u_j\rangle \delta_{ik}) - \nu \, [\langle u_i \, (\nabla_k u_j) \rangle+\langle u_j \, (\nabla_k u_i) \rangle]$
in Eq.~(\ref{C15}) describes the flux of $\tau_{ij}$.
Different effects related to budget equations~(\ref{C1})--(\ref{C15})
in a stratified turbulence have been discussed in a number of publications
\cite{CCH08,ZKR07,ZKR08,ZKR09,ZKR10,ZKR13,KRZ19,KRZ21,RKZ22}.

\subsection{Basic assumptions}
\label{subsec:3.2}

In the framework of the energy and flux budget turbulence closure theory, we assume
the following.
The characteristic times of variations of the densities of
the TKE, the potential temperature fluctuations intensity,
the turbulent flux of potential temperature and
the Reynolds stress
are substantially longer than the turbulent timescales.
This assumption yields the steady-state solutions
of the budget equations~(\ref{C1})--(\ref{C15}).

This allows one to express the dissipation rates of $E_{\rm K}$, $E_\theta$ and $F_i$ applying
the Kolmogorov  hypothesis. This implies that
$\varepsilon_{_{\rm K}}=E_{\rm K}/t_{\rm T}$, $\varepsilon_\theta=E_\theta/(C_{\rm p} \, t_{\rm T})$,
and $\varepsilon_i^{({\rm F})}=F_i/(C_{\rm F} \, t_{\rm T})$, where
$t_{\rm T}= \ell_z /E_{z}^{1/2}$ is the turbulent dissipation timescale, $\ell_z$ is the vertical integral scale,
and $C_{\rm p}$  and $C_{\rm F}$ are dimensionless
empirical constants.
In addition,  the dissipation rates $\varepsilon_{\alpha}$ of the TKE components $E_\alpha=\tau_{\alpha\alpha}$  with $\alpha=x, y, z$
are $\varepsilon_{x} = \varepsilon_{y}= \varepsilon_{z} =E_{\rm K}/3t_{\rm T}$  (see Ref.~\cite{LPR09}), since
the dominant contribution to $E_\alpha$ is from the Kolmogorov viscous scale
where turbulence is nearly isotropic.
Here the summation convention for the double Greek indices is not applied.

The term $\varepsilon_{i}^{(\tau)} = \varepsilon_{iz}^{(\tau)} - \beta \, F_i - Q_{iz}$ in Eq.~(\ref{C15})
is the effective dissipation rate of the off-diagonal components of the Reynolds stress $\tau_{iz}$  \cite{ZKR07,ZKR13,KRZ21}.
The dissipation rate of $\tau_{iz}$ is assumed to be due to
the combination $\varepsilon_{iz}^{(\tau)}- \beta \, F_i - Q_{iz}$, where $\varepsilon_{iz}^{(\tau)}
=\tau_{iz}/(C_\tau \, t_{\rm T})$.
Here $C_\tau$ is the effective-dissipation timescale empirical constant
\citep{ZKR07,ZKR13,KRZ21,RKZ22}.

The effective dissipation assumption was justified by Large Eddy Simulations (see Fig.~1 in Ref.~\cite{ZKR13}),
where the data from Ref.~\cite{ES04,EZ06}  were used
for the two types of atmospheric boundary layer: “nocturnal stable”
(with essentially negative buoyancy flux at the surface and neutral stratification in
the free flow) and “conventionally neutral” (with a negligible buoyancy flux at the
surface and essentially stably stratified turbulence in the free flow).
The effective dissipation assumption directly yields
the well-known down-gradient formulation of the vertical turbulent flux of momentum
$\langle u_i u_z \rangle = - K_{\rm M} \, \nabla_z \meanU_i $,
where $i=x, y$, and
\begin{eqnarray}
K_{\rm M} = 2 C_\tau \, t_{\rm T} \, E_z  = 2 C_\tau \, \ell_z \, E_z^{1/2}.
\label{LP2}
\end{eqnarray}
The latter result is valid for a shear-produced turbulence or a convective turbulence
with a non-uniform large-scale velocity field.
We point out that the diagonal components of the Reynolds stress
are much larger than the off-diagonal components.
The diagonal components of the Reynolds stress
determines the TKE components
which obey  the Kolmogorov spectrum $\propto k^{-5/3}$,
while the off-diagonal components of the Reynolds stress are produced by the tangling mechanism
of generation of anisotropic velocity fluctuations, and they obey the $\propto k^{-7/3}$ spectrum
\cite{IY02}.

The final assumption is related to the term $\rho_0^{-1} \, \langle \theta \,
\nabla_z p \rangle$ in the budget equation for  $F_z$ which is parameterized as
$\beta \, \langle \theta^2\rangle  - \rho_0^{-1} \, \langle \theta \,
\nabla_z p \rangle = 2 C_\theta \, \beta \, E_\theta $, where
$C_\theta <1$ is the positive dimensionless empirical constant.
The latter assumption has been justified analytically
(see Appendix~A in \cite{ZKR07}) and by Large Eddy Simulations,
where the data from Refs.~\cite{ES04,EZ06}  have been used for the two types
of atmospheric boundary layer: “nocturnal stable” and “conventionally neutral” (see Fig.~2 in \cite{ZKR13}).

For convective turbulence, we choose the following values of the non-dimensional empirical constants
which have been used in the EFB theory for stably stratified turbulence
\cite{KRZ21,RKZ22}: $C_{\rm p} = 0.417$, $C_\theta = 0.744$,
$C_\tau=0.1$ and $C_{\rm F}= 0.125$. This corresponds to the turbulent Prandtl number for
a non-stratified turbulence ${\rm Pr}_{_{\rm T}}^{(0)}=0.8$.

\subsection{Global characteristics of convective turbulence with semi-organized structures}
\label{subsec:3.3}

In this section we determine global characteristics of convective turbulence by averaging over the
entire volume of the semi-organized structure.
As follows from laboratory experiments \cite{NSS01,BEKR20},
direct numerical simulations \cite{BR16,KRB17,KA19,TSW17},
and mean-field numerical simulations \cite{OKR22},
the vertical gradient of the mean potential temperature
can be positive and negative inside the large-scale circulations in a convective turbulence.
In particular, the vertical gradient of the mean potential temperature
can be positive when the vertical turbulent flux of potential temperature
is negative.

To describe this effect, we introduce the degree of thermal anisotropy $\alpha$
in convective turbulence that depends on the form of plumes.
In particular, the plumes can be characterized by the two-point instantaneous correlation function
$\langle \theta(t,{\bm x}) \, u_z(t,{\bm x}+{\bm r}) \rangle$, where
$\ell_{\rm h}^{\rm (pl)} $ and $\ell_{z}^{\rm (pl)} $ are the horizontal and vertical
scales in which the two-point instantaneous correlation functions
$\langle \theta(t,{\bm x}) \, u_z(t,{\bm x}+{\bm r}) \rangle $ tend to 0
in the horizontal and vertical directions, respectively.
The degree of thermal anisotropy $\alpha$ is defined as  \cite{EKRZ02}
\begin{eqnarray}
\alpha = -3 \, \left({3- 4  \left(\ell_{\rm h}^{\rm (pl)} / \ell_{z}^{\rm (pl)}\right)^{2 \over 3} \over 2 + \left(\ell_{\rm h}^{\rm (pl)} / \ell_{z}^{\rm (pl)}\right)^{2 \over 3} }\right) .
\label{TD6}
\end{eqnarray}
For the isotropic case, $\ell_{\rm h}^{\rm (pl)}=\ell_{z}^{\rm (pl)}$
when the plumes have the form of ball and
the degree of thermal anisotropy $\alpha = 1$.
For $ \alpha < 1 $, the plumes are extended in the vertical direction having the form of
columns, $ \ell_{\rm h}^{\rm (pl)} < \ell_{z}^{\rm (pl)}$.
For $\ell_{\rm h}^{\rm (pl)} \ll \ell_{z}^{\rm (pl)}$, the parameter $ \alpha$ can be estimated as
\begin{eqnarray}
\alpha \approx  - {9 \over 2} \, \left(1 - {3 \over 2} \left({\ell_{\rm h}^{\rm (pl)} \over \ell_{z}^{\rm (pl)}}\right)^{2 \over 3}\right) .
\label{TD6a}
\end{eqnarray}
For $\alpha >1 $, the plumes have the form of  ``pancake'',
$ \ell_{\rm h}^{\rm (pl)} >\ell_{z}^{\rm (pl)}$.
For $\ell_{\rm h}^{\rm (pl)} \gg \ell_{z}^{\rm (pl)}$, the parameter $ \alpha$ can be estimated as
\begin{eqnarray}
\alpha \approx 12 \, \left(1 - {11 \over 4} \left({\ell_{z}^{\rm (pl)} \over \ell_{\rm h}^{\rm (pl)}}\right)^{2 \over 3}\right).
\label{TD6b}
\end{eqnarray}

Using  Eqs.~(\ref{BB12}) and~(\ref{LP2}), we determine
the normalized vertical turbulent flux of potential temperature averaged over the entire volume of the semi-organized structure as
\begin{eqnarray}
\langle \hat F \rangle_{_{V}} \equiv {\beta \, \langle F_z \rangle_{_{V}} \, \ell \over \langle E_{\rm K}^{3/2} \rangle_{_{V}}}
= C_\tau^2 \,\left({\ell_z \over L_z}\right)^2  \, {f_{\hat F}(A_\ast) \over {\rm Pr}_{_{T}} } ,
\label{TP10}
\end{eqnarray}
where $\langle ... \rangle_{_{V}}$ denote averaging over the entire volume of the semi-organized structure,
$\ell$ is the integral scale of the turbulence,
$\hat F=\beta \, F_z \, \ell/ E_{\rm K}^{3/2}$,
function $f_{\hat F}(A_\ast)$ is given by Eq.~~(\ref{AP2}) in Appendix~\ref{Appendix:A}.
We remind that the aspect ratio of the semi-organized structures is $A_\ast=\pi \, R / (\lambda \,  L_z)$
(see Sec.~\ref{sec:2}).
Here we assume that the turbulent dissipation timescale is
$t_{\rm T}= \ell_z /E_{z}^{1/2} =\ell /E_{\rm K}^{1/2}$
and the vertical anisotropy parameter is $A_z =E_z /E_{\rm K} = (\ell_z/\ell)^2 = 1/3$.

\begin{figure}
\vspace*{1mm}
\centering
\includegraphics[width=9.0cm]{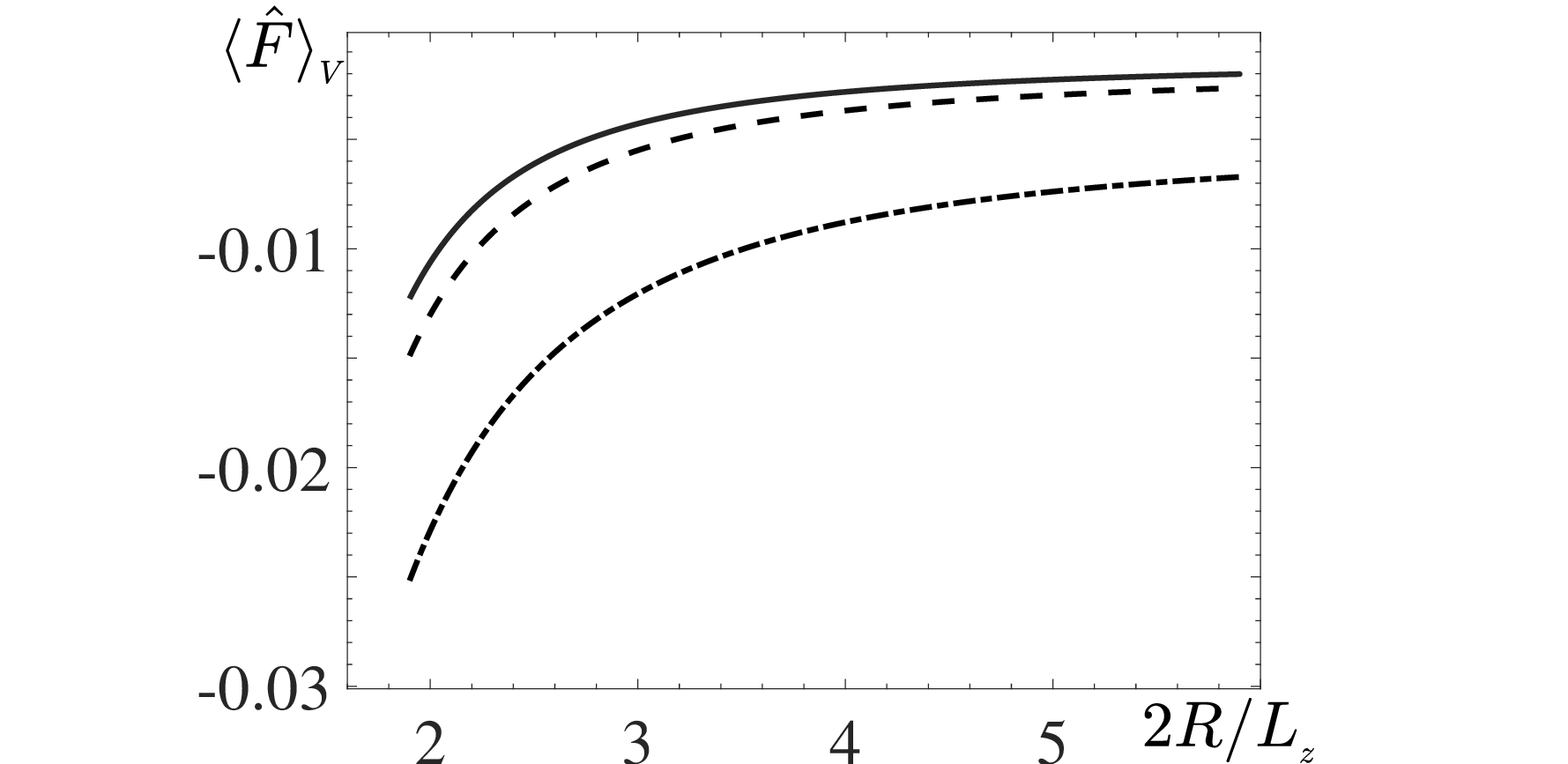}
\caption{\label{Fig3} The normalized vertical turbulent flux of the potential temperature
$\langle \hat F \rangle_{_{V}}$ versus the aspect ratio $2 R/L_z$ of the semi-organized structures
with the scale separation parameter $L_z/\ell_z=7$  and for different values of the  degree of thermal anisotropy:
$\alpha=-4$, i.e., for  $\ell_{\rm h}^{\rm (pl)} /\ell_{z}^{\rm (pl)} = 0.1$ (solid line); $\alpha=-2.1$, i.e., for  $\ell_{\rm h}^{\rm (pl)} /\ell_{z}^{\rm (pl)} = 0.2$ (dashed line) and $\alpha=-0.55$, i.e., for  $\ell_{\rm h}^{\rm (pl)} /\ell_{z}^{\rm (pl)} = 0.5$ (dashed-dotted line).
}
\end{figure}

As follows from Eqs.~(\ref{TP10}) and Eq.~~(\ref{AP2}) in Appendix~\ref{Appendix:A},
the vertical flux of potential temperature $\langle F_z \rangle_{_{V}}$ is negative when $f_{\hat F}(A_\ast)<0$, i.e.,
$2 \alpha \, \left(4 A_\ast^2 - 1\right) < 3 \left(1 + A_\ast^2\right)$.
This implies that the vertical flux of potential temperature $\langle F_z \rangle_{_{V}}$ is negative
when
\begin{eqnarray}
- {9 \over 2} < \alpha  < {3 \left(1 + A_\ast^2\right) \over 2 \, \left(4 A_\ast^2 - 1\right)} ,
\label{TD1a}
\end{eqnarray}
where we use Eq.~(\ref{TD6a}).
Note that the large-scale circulations are formed when $A_\ast \geq 1$ \cite{EKRZ02,EKRZ06}.
Applying Eqs.~(\ref{TD6}) and~(\ref{TD1a}), we obtain that
the vertical flux of potential temperature $\langle F_z \rangle_{_{V}}$ is negative when
\begin{eqnarray}
\left({\ell_{\rm h}^{\rm (pl)} \over \ell_{z}^{\rm (pl)}}\right)^{2 \over 3}  < {2\,  \left(13 A_\ast^2 -2\right) \over 31 A_\ast^2 - 9} .
\label{TD1b}
\end{eqnarray}
In Fig.~\ref{Fig3}, we plot the normalized vertical turbulent flux of the potential
temperature $\langle \hat F \rangle_{_{V}} \equiv \beta \, \langle F_z \rangle_{_{V}}
\, \ell / \langle E_{\rm K}^{3/2} \rangle_{_{V}}$ versus the aspect ratio $2 R/L_z$ of the semi-organized structures
for different values of the  degree of thermal anisotropy $\alpha$.
Here we take into account that semi-organized structures are formed when the scale separation parameter is $L_z/\ell_z > 5$
and the aspect ratio for the semi-organized structures is $2 R/L_z \geq 2$ \cite{EKRZ02,EKRZ06}.
It is seen in Fig.~\ref{Fig3} that when plumes are extended in the vertical direction, i.e.,
$\ell_{\rm h}^{\rm (pl)} /\ell_{z}^{\rm (pl)} < 1$, the vertical turbulent flux of
potential temperature $\langle F_z \rangle_{_{V}}$ is negative.

\begin{figure}
\vspace*{1mm}
\centering
\includegraphics[width=9.5cm]{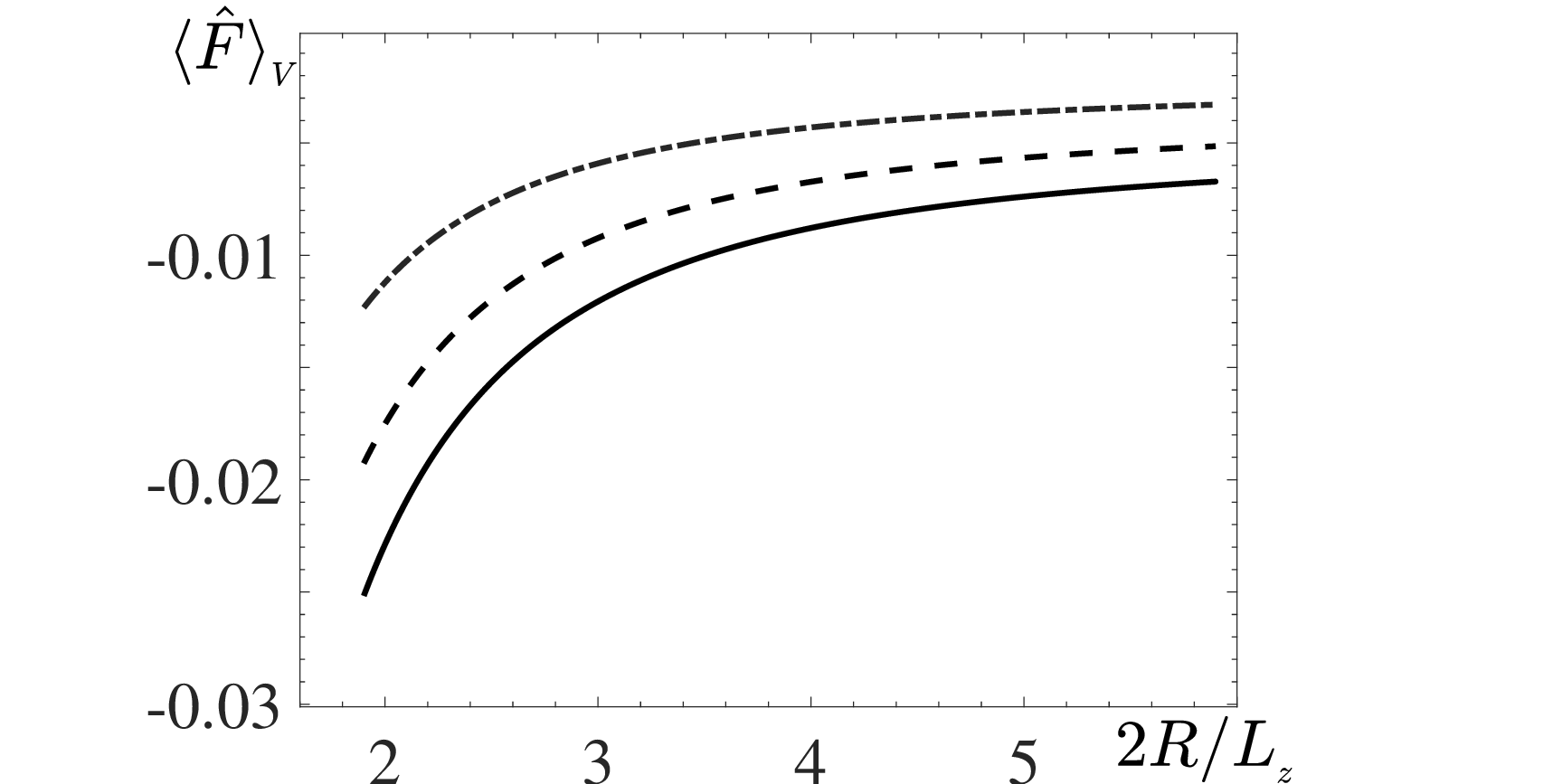}
\caption{\label{Fig4} The normalized vertical turbulent flux of the potential temperature $\langle \hat F \rangle_{_{V}} \equiv \beta \, \langle F_z \rangle_{_{V}} \, \ell / \langle E_{\rm K}^{3/2} \rangle_{_{V}}$ versus the aspect ratio $2 R/L_z$ of the semi-organized structures for $\alpha=-0.55$ and different values of the scale separation parameter $L_z/\ell_z=$
$7$ (solid line);  $8$ (dashed line) and $10$ (dashed-dotted line).
}
\end{figure}

In Fig.~\ref{Fig4}, we also show the dependence of
the normalized vertical turbulent flux of the potential temperature
$\langle \hat F \rangle_{_{V}}$ on the aspect ratio $2 R/L_z$ of the semi-organized structures
for $\alpha=-0.55$ ($\ell_{\rm h}^{\rm (pl)} /\ell_{z}^{\rm (pl)} = 0.5$)
and different values of the scale separation parameter $L_z/\ell_z$.
The absolute value of the normalized vertical turbulent flux of potential temperature decreases with
increase of scale separation between vertical size of the semi-organized structures $L_z$
and the vertical integral scale $\ell_z$.

As follows from Eq.~(\ref{TP10}) and Figs.~\ref{Fig2}--\ref{Fig3}, that the normalized vertical turbulent flux of potential temperature
averaged over the entire volume of the semi-organized structure is small
$(\langle \hat F \rangle_{_{V}} \ll 1)$ because the vertical integral scale $\ell_z$ is much smaller than
the vertical size $L_z$ of the semi-organized structure. Note also the coefficient $C_\tau$ is small.
This means that the volume averaged TKE dissipation rate $\langle \varepsilon_{_{\rm K}} \rangle_{_{V}} \sim \langle E_{\rm K} \rangle_{_{V}}/t_{\rm T} \sim \langle E_{\rm K}^{3/2} \rangle_{_{V}}/\ell$, is much larger than the turbulence production rate,  $\beta \, \langle F_z \rangle_{_{V}}$, caused by buoyancy.
This effect is due to the fact that for large Rayleigh numbers, the convective  turbulence
is mainly produced by the local shear of the semi-organized structures rather than the buoyancy (see below).

The production rate, $\Pi_{\rm K}$, of the turbulent kinetic energy by local large-scale shear $\nabla_j \meanU_i^{({\rm s})}$
of the semi-organized structures is given by $\Pi_{\rm K} = - \tau_{ij} \, \nabla_j \meanU_i^{({\rm s})} = 2 K_{\rm M} S^2$
[see the first term on the RHS of Eq.~(\ref{C1})],
where $S$ is the large-scale shear.
Using the steady-state version of the budget equation~(\ref{C1}), we obtain that
turbulent kinetic energy is $E_{\rm K}=4 C_\tau \, \ell_z^2 S^2/(1- \hat F)$.
To find the production rate of the turbulent kinetic energy
averaged over the entire volume of the semi-organized structure, we use the analytical
solution~(\ref{BB6})--(\ref{BB7}) for the velocity $\meanUU^{({\rm s})}$ of the semi-organized structure.
This yields the averaged squared large-scale shear $\left\langle S^2\right\rangle_{_{V}}$
given by Eq.~(\ref{TP1}) in Appendix~\ref{Appendix:B}.
Therefore, the turbulent kinetic energy density
averaged over the entire volume of the semi-organized structure, is given by
\begin{eqnarray}
\langle E_{\rm K} \rangle_{_{V}} = C_\tau  \, \left({\ell_z \over L_z}\right)^2 \, \meanU_{z0}^2 \, {f_S(A_\ast) \over 1 - \langle \hat F \rangle_{_{V}}},
\label{TP9}
\end{eqnarray}
where the function $f_S(A_\ast)$ is given by Eq.~(\ref{AP1}) in Appendix~\ref{Appendix:A}.
Equation~(\ref{TP9}) implies that the turbulent kinetic energy density $\langle E_{\rm K} \rangle_{_{V}}$
is much less than the squared velocity $\meanU_{z0}^2$.

The vertical flux of potential temperature, $\left\langle \overline{\Theta}^{({\rm s})} \meanU_{z}^{({\rm s})} \right\rangle_{_{V}}$, transported by the semi-organized structures, is determined using Eqs.~(\ref{BB11}) and (\ref{TP9}),
and Eq.~(\ref{TTTP12})  in Appendix~\ref{Appendix:B}:
\begin{eqnarray}
&& \left\langle \overline{\Theta}^{({\rm s})} \meanU_{z}^{({\rm s})} \right\rangle_{_{V}}
= C_\tau^{3/2} \, \left({\ell_z^2 \, \meanU_{z0}^3 \over L_z^3 \, \beta}\right)
\, {f_{F_{s}}(A_\ast) \over  \left(1 - \langle \hat F \rangle_{_{V}}\right)^{1/2} } ,
\label{TTP12}
\end{eqnarray}
where function $f_{F_{s}}(A_\ast)$ is given by Eq.~(\ref{AP4}) in Appendix~\ref{Appendix:A}.
By means of Eq.~(\ref{TTP12}), we find a characteristic convective velocity  $U_{\rm D}$ defined as
\begin{eqnarray}
U_{\rm D} &\equiv& \left(\beta \left\langle \overline{\Theta}^{({\rm s})}
\meanU_{z}^{({\rm s})} \right\rangle_{_{V}} L_z\right)^{1 \over 3}
\nonumber\\
&=& \meanU_{z0} \, \, C_\tau^{1/2} \,   \left({\ell_z \over L_z}\right)^{2 \over 3} \,
\, {f_{F_{s}}^{1/3}(A_\ast) \over  \left(1 - \langle \hat F \rangle_{_{V}}\right)^{1/6} } .
\label{TTP14}
\end{eqnarray}
Now we define a characteristic convective temperature $\Theta_{\rm D}$
from a condition
\begin{eqnarray}
\Theta_{\rm D} \, U_{\rm D} = \left\langle \overline{\Theta}^{({\rm s})} \meanU_{z}^{({\rm s})}
\right\rangle_{_{V}} .
\label{TDIF14}
\end{eqnarray}
Using the definition~(\ref{TTP14}) of the convective velocity  $U_{\rm D}$,
we obtain a relation between the convective velocity $U_{\rm D}$ and the convective temperature $\Theta_{\rm D}$ as
$U_{\rm D}=(\beta L_z \Theta_{\rm D})^{1/2}$.
By means of Eq.~(\ref{TTP14})
and Eq.~(\ref{TTTP12})  in Appendix~\ref{Appendix:B},
we find the convective temperature $\Theta_{\rm D}$ as
\begin{eqnarray}
\Theta_{\rm D} = \overline{\Theta}_{0} \,  \left({\ell_z \over L_z}\right)^{-{2\over 3}}
\, {\left(1 - \langle \hat F \rangle_{_{V}}\right)^{1/6}  \over 2 C_\tau^{1/2} \, f_{F_{s}}^{1/3}(A_\ast)} \, J_0^2 (\lambda) .
\label{TTTP27}
\end{eqnarray}
The velocity $U_{\rm D}$ and temperature $\Theta_{\rm D}$ characterize
the large-scale properties of convection.

Let us determine the global energetic characteristics of semi-organized structures.
Equations~(\ref{TTP14})
and~(\ref{TTTP1})  in Appendix~\ref{Appendix:B}
yield an expression for the kinetic energy density of semi-organized structures as
\begin{eqnarray}
\langle  E_{\rm U} \rangle_{_{V}} &\equiv& { 1 \over 2} \, \left[\left\langle \left(\meanU_{z}^{({\rm s})}\right)^2 \right\rangle_{_{V}}
+ \left\langle \left(\meanU_{r}^{({\rm s})}\right)^2 \right\rangle_{_{V}} \right]
 \nonumber\\
&=&{U_{\rm D}^2 \over C_\tau} \, \left({L_z \over \ell_z}\right)^{4 \over 3} \,
\,\left(1 - \langle \hat F \rangle_{_{V}}\right)^{1 \over 3} \, f_U(A_\ast) ,
\label{TP23}
\end{eqnarray}
where function $f_U(A_\ast)$ is given by Eq.~(\ref{AP5}) in Appendix~\ref{Appendix:A}.

In Fig.~\ref{Fig5}, we show the normalized velocity $\tilde U = \sqrt{2 \langle  E_{\rm U} \rangle_{_{V}}}/U_{\rm D}$
versus the aspect ratio $2 R/L_z$ of the semi-organized structures
for different values of the scale separation parameter $L_z/\ell_z$.
The kinetic energy density $\langle  E_{\rm U} \rangle_{_{V}}$ of the semi-organized structures
increases with increase of the scale separation parameter $L_z/\ell_z$ [see Eq.~(\ref{TP23})].
Note that the kinetic energy density $\langle  E_{\rm U} \rangle_{_{V}}$ is nearly independent of the parameter
 $\alpha$ (and the ratio $\ell_{\rm h}^{\rm (pl)} /\ell_{z}^{\rm (pl)}$) which characterizes
 the thermal anisotropy of convective turbulence.

\begin{figure}
\vspace*{1mm}
\centering
\includegraphics[width=9.5cm]{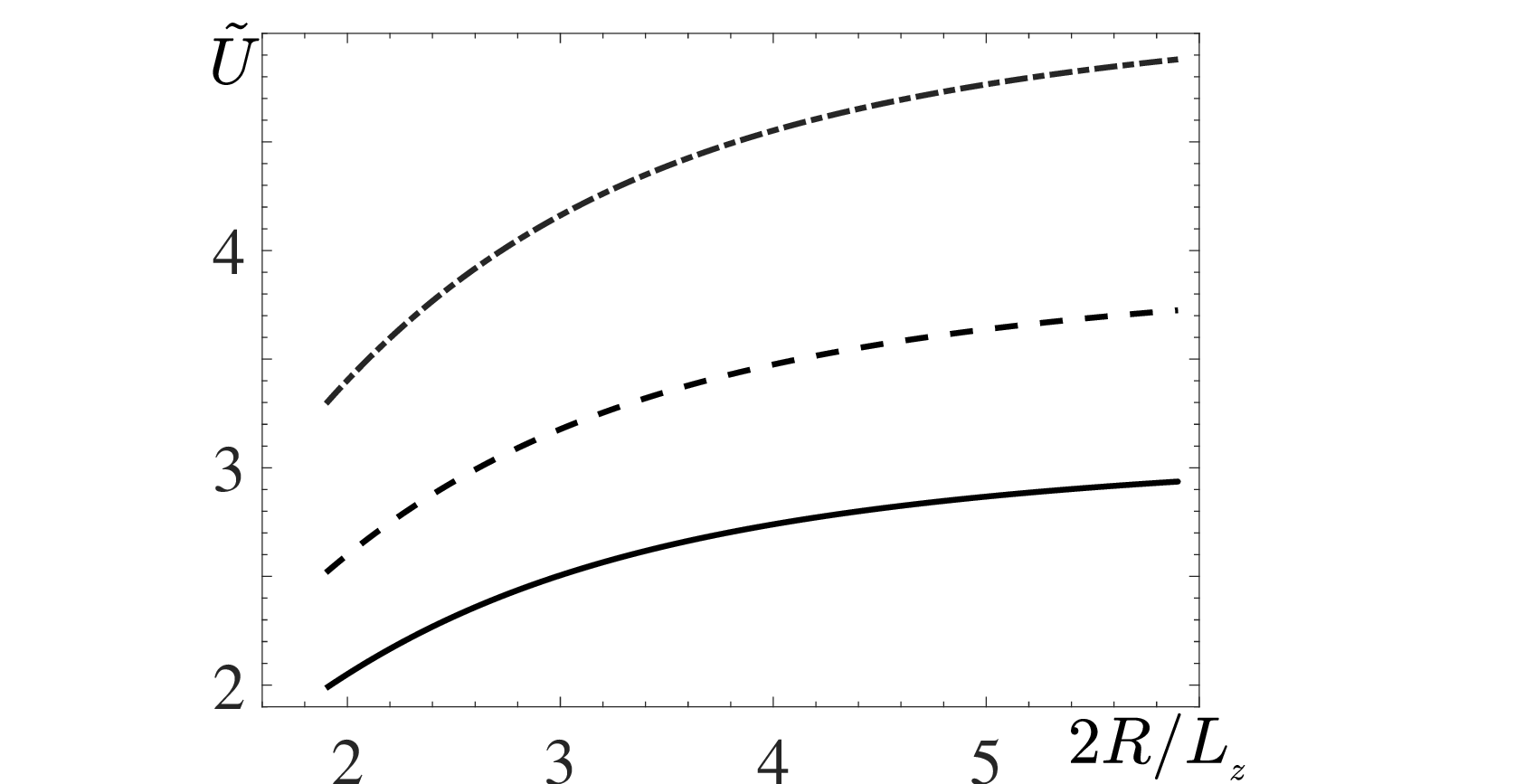}
\caption{\label{Fig5} The normalized velocity $\tilde U = \sqrt{2 \langle  E_{\rm U} \rangle_{_{V}}}/U_{\rm D}$
versus the aspect ratio $2 R/L_z$ of the semi-organized structures
for $\alpha=-0.55$ and different values of the scale separation parameter $L_z/\ell_z=$
$7$ (solid line);  $8$ (dashed line) and $10$ (dashed-dotted line).
}
\end{figure}

Equation~(\ref{TTTP27})
and Eq.~(\ref{BP6})  in Appendix~\ref{Appendix:B}
yield the thermal energy density of the semi-organized structures as
\begin{eqnarray}
\langle E_\Theta \rangle_{_{V}} &\equiv& { 1 \over 2} \, \left\langle \left(\overline{\Theta}^{({\rm s})}\right)^2 \right\rangle_{_{V}}
\nonumber\\
&=& \Theta_{\rm D}^2 \, C_\tau  \, \left({\ell_z \over L_z}\right)^{4\over 3}  \,
{f_\Theta(A_\ast)  \over  \left(1 - \langle \hat F \rangle_{_{V}}\right)^{1/3}} ,
\label{TP27}
\end{eqnarray}
where function $f_\Theta(A_\ast)$ is given by Eq.~(\ref{AP6}) in Appendix~\ref{Appendix:A}.
In Fig.~\ref{Fig6}, we plot the normalized potential temperature
$\tilde \Theta = \sqrt{2 \langle  E_\Theta \rangle_{_{V}}}/\Theta_{\rm D}$
versus the aspect ratio $2 R/L_z$ of the semi-organized structures
for different values of the scale separation parameter $L_z/\ell_z$.
Equation~(\ref{TP27})  and Fig.~\ref{Fig6} demonstrate that
the thermal energy  density of the semi-organized structure $\langle  E_\Theta \rangle_{_{V}}$
increases with the aspect ratio $2 R/L_z$ of the semi-organized structures
approaching to the value which is of the order of $\Theta_{\rm D}^2$.
Equations~(\ref{TP23}) and~(\ref{TP27}) imply that the flux $\tilde U \, \tilde \Theta$
of the potential temperature transported by the semi-organized structures
is independent of the scale separation parameter $L_z/\ell_z$.

\begin{figure}
\vspace*{1mm}
\centering
\includegraphics[width=9.5cm]{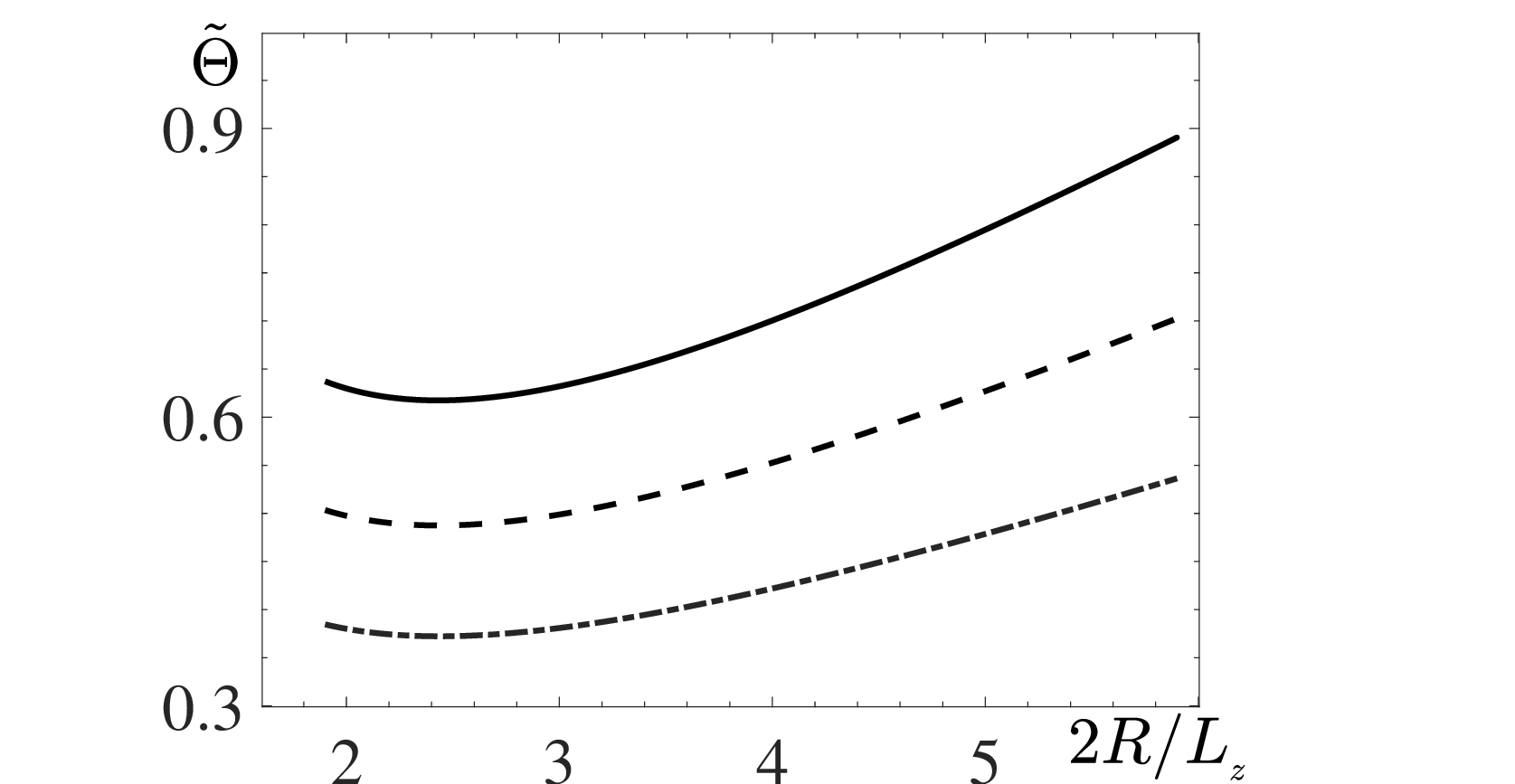}
\caption{\label{Fig6} The normalized temperature $\tilde \Theta = \sqrt{2 \langle  E_\Theta \rangle_{_{V}}}/\Theta_{\rm D}$
versus the aspect ratio $2 R/L_z$ of the semi-organized structures
for $\alpha=-0.55$ and different values of the scale separation parameter $L_z/\ell_z=$
$7$ (solid line);  $8$ (dashed line) and $10$ (dashed-dotted line).
}
\end{figure}

\begin{figure}
\vspace*{1mm}
\centering
\includegraphics[width=9.5cm]{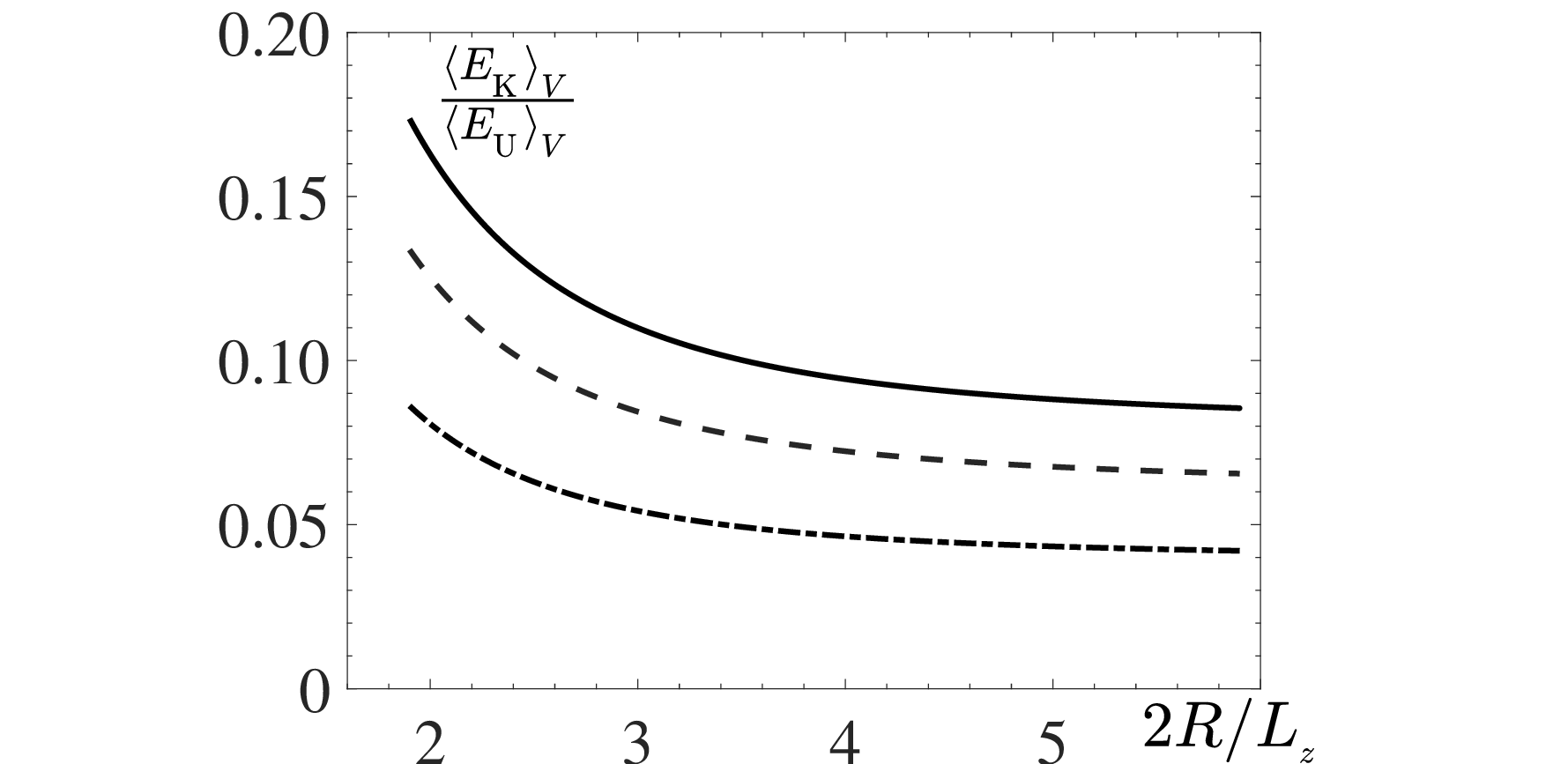}
\caption{\label{Fig7} The ratio of the turbulent kinetic energy  density to the kinetic energy density of semi-organized structures,
$\langle E_{\rm K} \rangle_{_{V}} /  \langle  E_{\rm U} \rangle_{_{V}}$
versus the aspect ratio $2 R/L_z$ of the semi-organized structures
for $\alpha=-0.55$ and different values of the scale separation parameter $L_z/\ell_z=$
$7$ (solid line);  $8$ (dashed line) and $10$ (dashed-dotted line).
}
\end{figure}

Using Eqs.~(\ref{TP9}) and~(\ref{TTP14}), we express the turbulent kinetic energy  density
$\langle E_{\rm K} \rangle_{_{V}}$ in terms of the squared convective velocity
$U_{\rm D}^2$ as
\begin{eqnarray}
\langle E_{\rm K} \rangle_{_{V}} &=& U_{\rm D}^2 \, \left({\ell_z \over L_z}\right)^{2\over 3} \, \, {f_u(A_\ast)
\over  \left(1 - \langle \hat F \rangle_{_{V}}\right)^{2/3} }  ,
\label{TP24}
\end{eqnarray}
where function $f_u(A_\ast)$ is given by Eq.~(\ref{AP7}) in Appendix~\ref{Appendix:A}.
Equation~(\ref{TP24})  implies that the turbulent kinetic energy density
is much smaller than the squared velocity $U_{\rm D}^2$.
Equations~(\ref{TP23}) and~(\ref{TP24}) allow to determine the ratio of the turbulent kinetic energy  density $\langle E_{\rm K} \rangle_{_{V}}$ to the kinetic energy density of semi-organized structures $\langle  E_{\rm U} \rangle_{_{V}}$
as
\begin{eqnarray}
{\langle E_{\rm K} \rangle_{_{V}} \over  \langle  E_{\rm U} \rangle_{_{V}}} =C_\tau \, \left({\ell_z \over L_z}\right)^{2}
\,  \left(1 - \langle \hat F \rangle_{_{V}}\right)^{-1} \, {f_u(A_\ast) \over f_U(A_\ast)} .
\label{TP28}
\end{eqnarray}
In Fig.~\ref{Fig7}, we plot the ratio of the turbulent kinetic energy  density to the kinetic energy density of semi-organized structures,
$\langle E_{\rm K} \rangle_{_{V}} /  \langle  E_{\rm U} \rangle_{_{V}}$
versus the aspect ratio $2 R/L_z$ of the semi-organized structures
for different values of the scale separation parameter $L_z/\ell_z$.
As follows from Eq.~(\ref{TP28}) and Fig.~\ref{Fig7},
the turbulent kinetic energy density $\langle E_{\rm K} \rangle_{_{V}}$
is much smaller than the kinetic energy density of semi-organized structures
$\langle  E_{\rm U} \rangle_{_{V}}$.
This is because the vertical integral scale $\ell_z$ is much smaller than
the vertical size $L_z$ of the semi-organized structure.
Indeed, intensity of velocity fluctuations can be estimated as $\langle {\bm u}^2 \rangle \sim t_{\rm T} K_{\rm M} S^2 \sim \left[\meanU_{r}^{({\rm s})} \ell_z /L_z\right]^2$, where
we take into account that the production rate of the turbulent kinetic energy density is due to the local large-scale shear
of the semi-organized structures $- \tau_{ij} \, \nabla_j \meanU_i^{({\rm s})} \sim K_{\rm M} S^2$
and the large-scale shear is estimated as $S \sim \meanU_{r}^{({\rm s})}/L_z$.

Equations~(\ref{TP10}), (\ref{TP9})  and~(\ref{TTP12}) allow us to determine the
ratio of the vertical turbulent flux of potential temperature $\langle F_z \rangle_{_{V}}$
to the vertical flux of potential temperature $\left\langle \overline{\Theta}^{({\rm s})} \meanU_{z}^{({\rm s})} \right\rangle_{_{V}}=
\Theta_{\rm D} \, U_{\rm D}$ transported by the semi-organized structures
 as
\begin{eqnarray}
{\langle F_z \rangle_{_{V}} \over  \Theta_{\rm D} \, U_{\rm D}}
= {C_\tau^{2} \over {\rm Pr}_{_{T}}} \,  \left({\ell_z \over L_z}\right)^2  \left(1 - \langle \hat F \rangle_{_{V}}\right)^{-1} \, f_{F_{T}}(A_\ast)  ,
\label{TP14}
\end{eqnarray}
where function $f_{F_{T}}(A_\ast)$ is given by Eq.~(\ref{AP3}) in Appendix~\ref{Appendix:A}.
In Fig.~\ref{Fig8} we show the normalized vertical turbulent flux of the potential temperature
$\langle F_z \rangle_{_{V}}/(\Theta_{\rm D} \, U_{\rm D})$ versus the aspect ratio $2 R/L_z$ of the
semi-organized structures for $\alpha=-0.55$ (i.e., for $\ell_{\rm h}^{\rm (pl)} /\ell_{z}^{\rm (pl)} = 0.5$)
and different values of the scale separation parameter $L_z/\ell_z$.
It follows from Eq.~(\ref{TP14}) and Fig.~\ref{Fig8}
that the vertical flux of potential temperature,
$\left\langle \overline{\Theta}^{({\rm s})} \meanU_{z}^{({\rm s})} \right\rangle_{_{V}}$, transported by the semi-organized structures is much larger than the vertical turbulent flux,  $\langle F_z \rangle_{_{V}}$,  of potential temperature, i.e.,
$|\langle F_z \rangle_{_{V}}|/(\Theta_{\rm D} \, U_{\rm D}) \ll 1$.
This is because the vertical integral scale $\ell_z$ is much smaller than
the vertical size $L_z$ of the semi-organized structure.
Indeed, the vertical turbulent flux can be estimated as $|F_z| \sim t_{\rm T} \langle u_z^2 \rangle \left(\nabla_z \overline{\Theta}^{({\rm s})} \right) \sim t_{\rm T}^2 \, K_{\rm M} S^2 \,  \left(\nabla_z \overline{\Theta}^{({\rm s})}\right) \sim (t_{\rm T} \,  S) \, \left[ \Theta_{\rm D} \, U_{\rm D} \, \ell_z /L_z\right]^2$, where we take into account that $\langle u_z^2 \rangle \sim t_{\rm T} K_{\rm M} S^2$,
the shear is estimated as $S \sim U_{\rm D}/L_z$, the gradient of the mean potential temperature is estimated as $\nabla_z \overline{\Theta}^{({\rm s})} \sim \Theta_{\rm D} /L_z$ and $t_{\rm T} \,  S \leq 1$.
Note that the analysis of large-scale instability in shear-free convection \cite{EKRZ02,EKRZ06}
shows that the semi-organized structures are formed when the scale separation parameter $L_z/\ell_z>5$.

\begin{figure}
\vspace*{1mm}
\centering
\includegraphics[width=9.5cm]{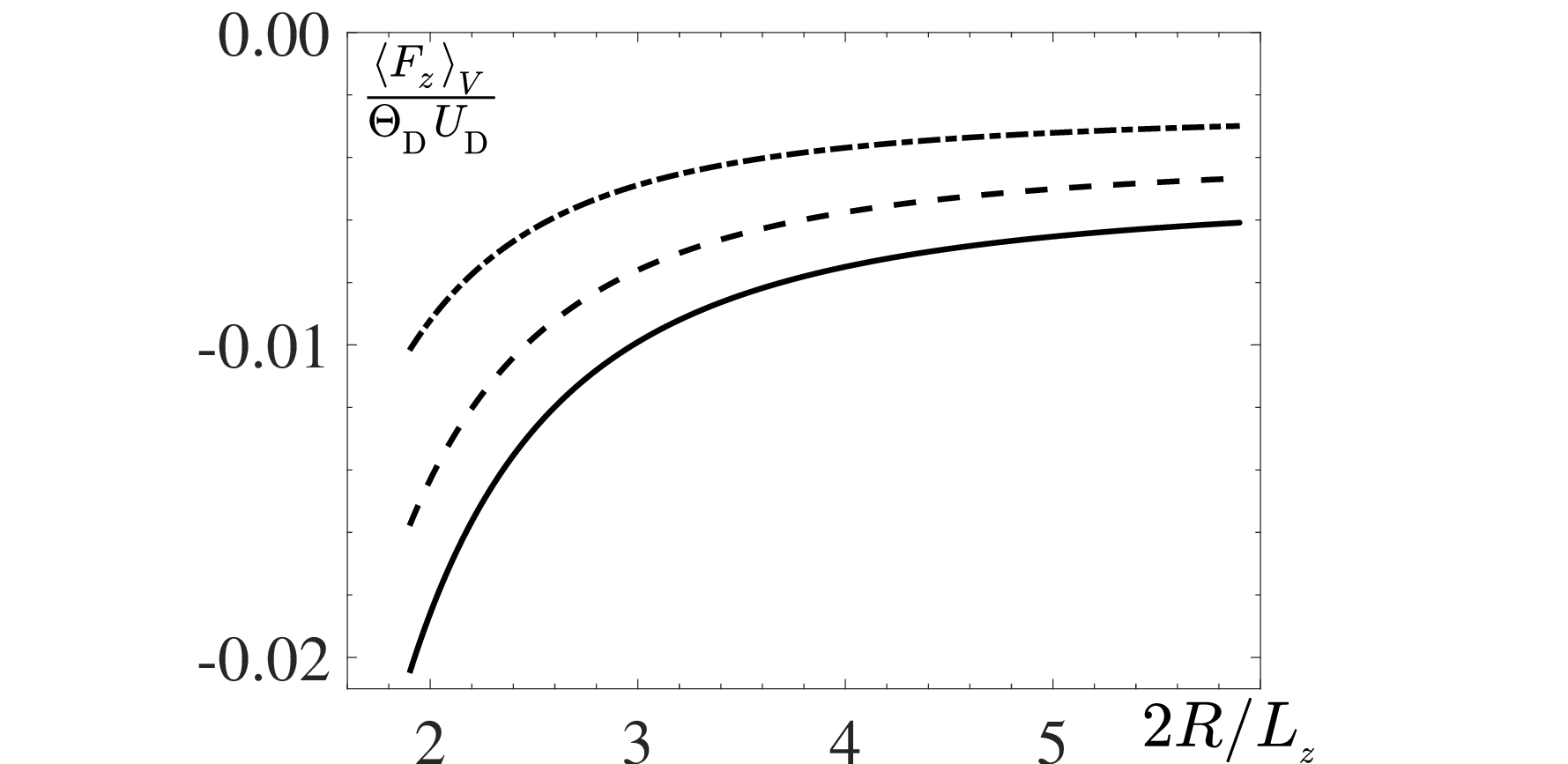}
\caption{\label{Fig8} The normalized vertical turbulent flux of the potential temperature
$\langle F_z \rangle_{_{V}}/(\Theta_{\rm D} \, U_{\rm D})$ versus the aspect ratio $2 R/L_z$ of the
semi-organized structures for $\alpha=-0.55$ (i.e., for $\ell_{\rm h}^{\rm (pl)} /\ell_{z}^{\rm (pl)} = 0.5$)
and different values of the scale separation parameter $L_z/\ell_z=$
$7$ (solid line);  $8$ (dashed line) and $10$ (dashed-dotted line).
}
\end{figure}

\begin{figure}
\vspace*{1mm}
\centering
\includegraphics[width=9.0cm]{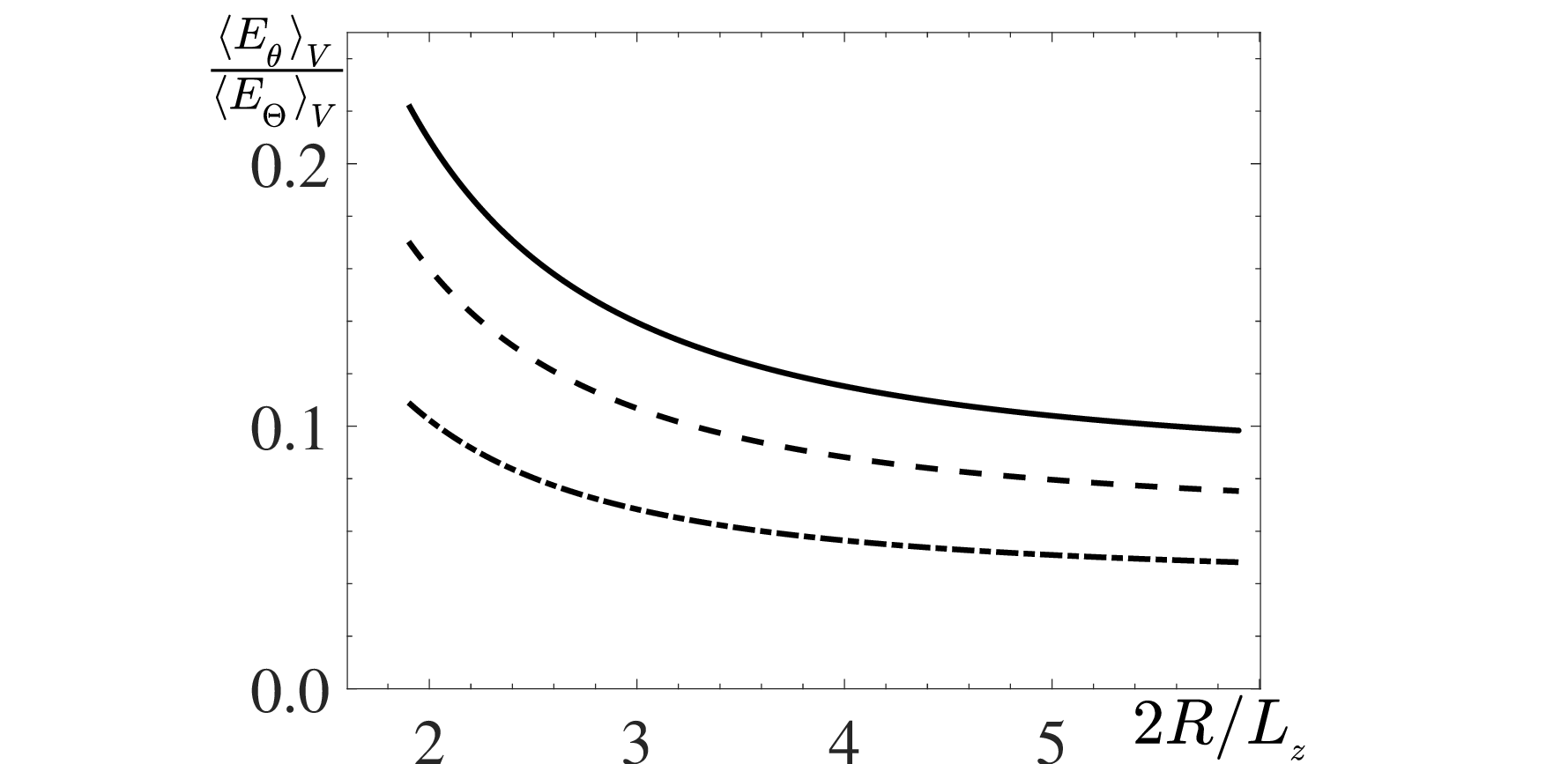}
\caption{\label{Fig9} The ratio of the turbulent thermal energy  density
to the thermal energy  density of the semi-organized structures, $\langle E_\theta \rangle_{_{V}}/\langle E_\Theta \rangle_{_{V}}$,
versus the aspect ratio $2 R/L_z$ of the semi-organized structures for $\alpha=-0.55$
and different values of the scale separation parameter $L_z/\ell_z=$
$7$ (solid line);  $8$ (dashed line) and $10$ (dashed-dotted line).
}
\end{figure}

Using Eqs.~(\ref{TP15})--(\ref{TTP18}) in Appendix~\ref{Appendix:B},
we obtain expression for the turbulent thermal energy density $\langle E_\theta \rangle_{_{V}}$ as
\begin{eqnarray}
\left\langle E_\theta \right\rangle_{_{V}} &=& \Theta_{\rm D}^2 \,  C_{\rm F} \, C_{\rm p} \, C_\tau \, \left({\ell_z \over L_z}\right)^{10\over 3} \, \, {f_\theta(A_\ast) \over  \left(1 - \langle \hat F \rangle_{_{V}}\right)^{1/3} }  ,
\nonumber\\
\label{TP21}
\end{eqnarray}
where function $f_\theta(A_\ast)$ is given by Eq.~(\ref{AP8}) in Appendix~\ref{Appendix:A}.
Thus, the ratio of the turbulent thermal energy  density $\langle E_\theta \rangle_{_{V}}$
to the thermal energy density of the semi-organized structures $\langle E_\Theta \rangle_{_{V}}$
is given by
\begin{eqnarray}
{\langle E_\theta \rangle_{_{V}} \over \langle E_\Theta \rangle_{_{V}}} = C_{\rm F} \, C_{\rm p} \, \left({\ell_z \over L_z}\right)^{2} \, {f_\theta(A_\ast)  \over f_\Theta(A_\ast)},
\label{TP29}
\end{eqnarray}
where we use Eq.~(\ref{TP27}).
In Fig.~\ref{Fig9} we show the ratio of the turbulent thermal energy  density
to the thermal energy  density of the semi-organized structures, $\langle E_\theta \rangle_{_{V}}/\langle E_\Theta \rangle_{_{V}}$,
versus the aspect ratio $2 R/L_z$ of the semi-organized structures for
different values of the scale separation parameter $L_z/\ell_z$.
Equation~(\ref{TP29}) and Fig.~\ref{Fig9}
demonstrate that the turbulent thermal energy $\langle E_\theta \rangle_{_{V}}$ is much smaller than
the thermal energy of the semi-organized structures $\langle E_\Theta \rangle_{_{V}}$.
This is because the vertical integral scale $\ell_z$ is much smaller than
the vertical size $L_z$ of the semi-organized structure.
Indeed, the turbulent thermal energy density can be estimated as $\langle \theta^2 \rangle \sim t_{\rm T} K_{\rm H} \left(\nabla_z \overline{\Theta}^{({\rm s})}\right)^2 \sim \left[\overline{\Theta}^{({\rm s})} \ell_z /L_z\right]^2$, where
we take into account that the production rate of the potential temperature fluctuations is
estimated as $ - ({\bm F}{\bf \cdot} \bec{\nabla})\overline{\Theta}^{({\rm s})} \sim K_{\rm H} \left(\nabla_z \overline{\Theta}^{({\rm s})}\right)^2$, the vertical turbulent flux of the potential temperature is $F_z \sim -K_{\rm H} \, \nabla_z \overline{\Theta}^{({\rm s})}$
and the gradient of the mean potential temperature is estimated as $\nabla_z \overline{\Theta}^{({\rm s})} \sim \overline{\Theta}^{({\rm s})} /L_z$.

Using Eqs.~(\ref{TP14}) and~(\ref{TP30})  in Appendix~\ref{Appendix:B},
we obtain the expression for the vertical gradient of the mean potential temperature,
$\nabla_z \overline{\Theta}^{({\rm m})} \approx - \langle F_z \rangle_{_{V}}/ \langle K_{\rm H} \rangle_{_{V}} $,
as
\begin{eqnarray}
\nabla_z \overline{\Theta}^{({\rm m})} = - {\Theta_{\rm D} \over L_z}\left({\ell_z \over L_z}\right)^{2\over 3} \,  \left(1 - \langle \hat F \rangle_{_{V}}\right)^{-4/3} \, f_{_{\nabla}}(A_\ast) ,
\nonumber\\
\label{TP31}
\end{eqnarray}
where function $f_{_{\nabla}}(A_\ast)$ is given by Eq.~(\ref{AP10}) in Appendix~\ref{Appendix:A}.
In Fig.~\ref{Fig10} we plot the normalized vertical gradient
of the mean potential temperature, $\nabla_z \tilde\Theta = (L_z / \Theta_{\rm D})\, \nabla_z \overline{\Theta}^{({\rm m})}$,
versus the aspect ratio $2 R/L_z$ of the semi-organized structures for different values of the scale separation parameter $L_z/\ell_z$.
As follows from Eq.~(\ref{TP31}) and Fig.~\ref{Fig10} that the normalized vertical gradient
of the mean potential temperature $\nabla_z \tilde\Theta$ is small and positive, because for the considered parameter range
[see Eq.~(\ref{TD1a})] the vertical turbulent flux of
the potential temperature  $\langle F_z \rangle_{_{V}}$ is negative as well as
the function $f_{_{\nabla}}(A_\ast)$ is negative.

\begin{figure}
\vspace*{1mm}
\centering
\includegraphics[width=9.0cm]{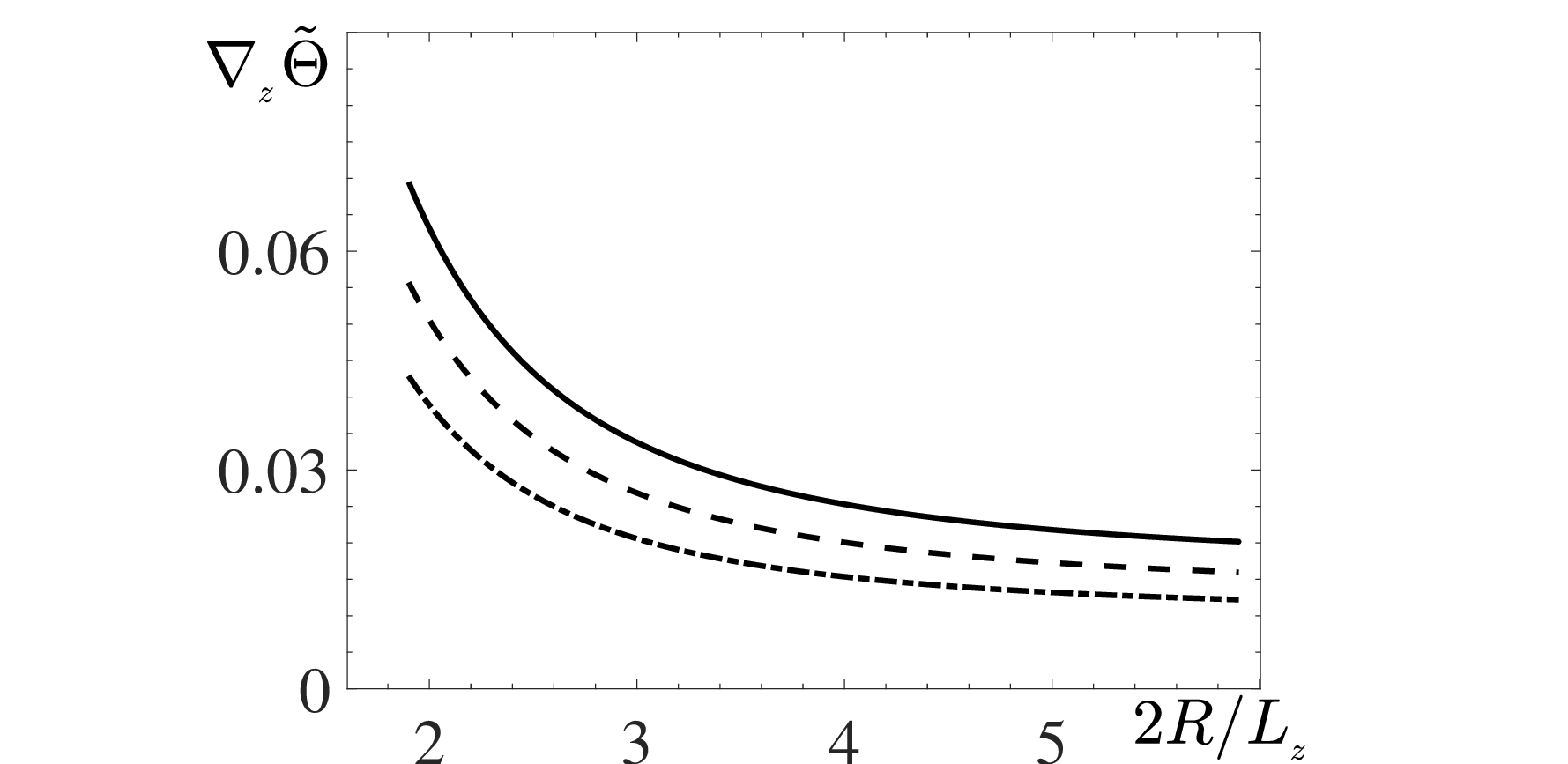}
\caption{\label{Fig10} The normalized vertical gradient of the mean potential temperature, $\nabla_z \tilde\Theta = (L_z / \Theta_{\rm D})\, \nabla_z \overline{\Theta}^{({\rm m})} $,
versus the aspect ratio $2 R/L_z$ of the semi-organized structures for $\alpha=-0.55$
and different values of the scale separation parameter $L_z/\ell_z=$
$7$ (solid line);  $8$ (dashed line) and $10$ (dashed-dotted line).
}
\end{figure}

Therefore, the global characteristics of a convective turbulence depend on the aspect ratio of the semi-organized structures,
the scale separation parameter between the vertical size $L_z$ of the structures and the integral scale of turbulence $\ell_z$,
and the degree of thermal anisotropy (i.e., the form of plumes).
In the limit of large aspect ratio of the semi-organized structures,
the global turbulence characteristics reaches their asymptotic values which depend
on the degree of thermal anisotropy and the scale separation parameter
(see Figs.~\ref{Fig2}--\ref{Fig3}, \ref{Fig6}--\ref{Fig9}).

 \section{Transition from CBL core to convective surface layer}
 \label{sec:4}

 In this Section we discuss a matching of solutions obtained for the CBL core and the convective surface layer.
 We start with the solutions obtained for the convective surface layer.

 \subsection{Convective surface layer}
\label{subsec:4.1}

\begin{figure}
\vspace*{1mm}
\centering
\includegraphics[width=9.5cm]{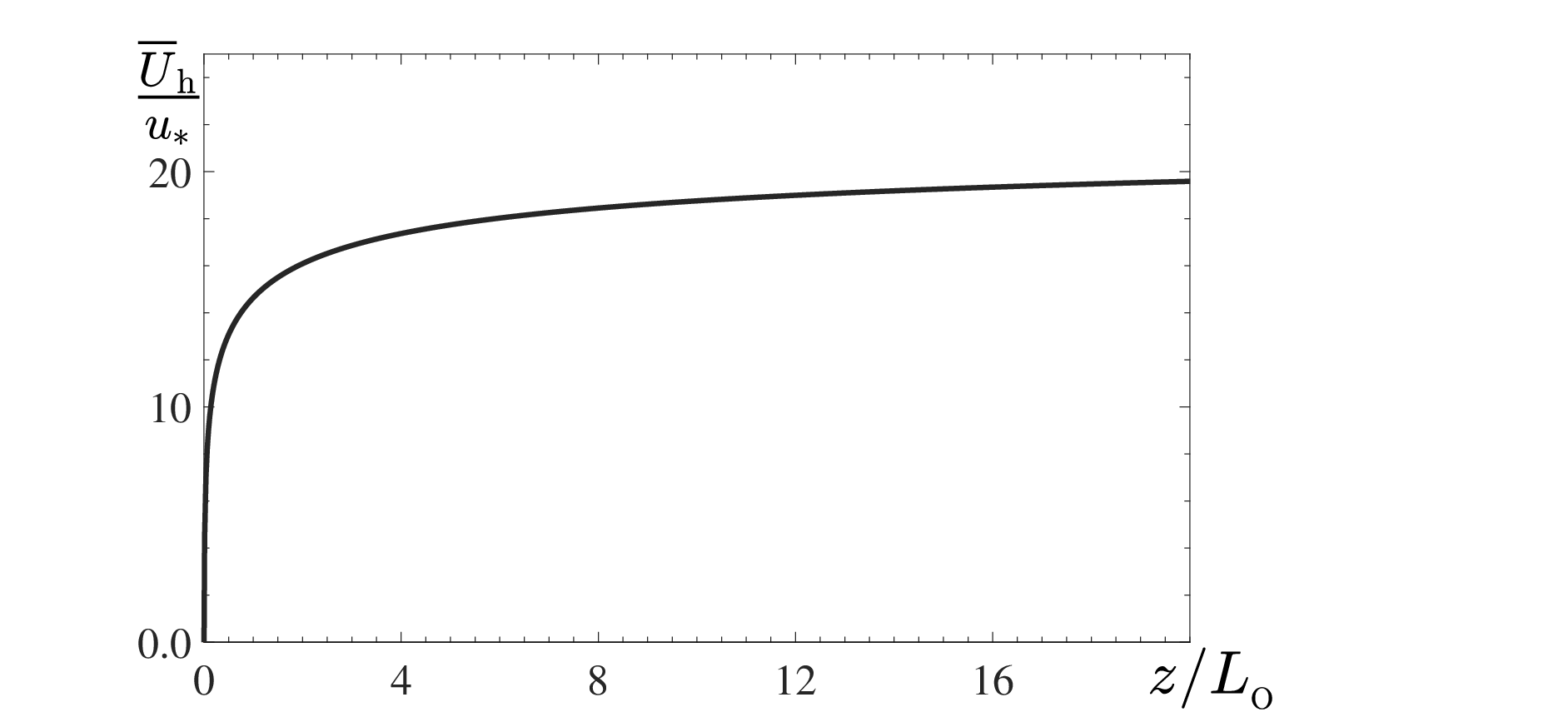}
\caption{\label{Fig11}
The vertical profile of the normalized mean
horizontal velocity  $\meanU_{\rm h}(z)/u_\ast$ at the surface layer.
}
\end{figure}

First, we outline the results of the EFB theory for the atmospheric
convective surface layer \cite{RKZ22}.
The vertical profiles of various turbulent characteristics
are determined by the following equations:
\begin{itemize}
\item{
The flux Richardson number,
\begin{eqnarray}
{\rm Ri}_{\rm f}(\tilde Z) = \tilde Z \, \left[\tilde E_{\rm K}(\tilde Z)\right]^{1/2} ,
\label{DG2}
\end{eqnarray}
where ${\rm Ri}_{\rm f} = -  |\tilde Z|$ for $ |\tilde Z|  \ll 1$ and ${\rm Ri}_{\rm f} = -  \tilde Z^{4/3}$ for $ |\tilde Z|  \gg 1$;
}
\item{
the turbulent viscosity,
\begin{eqnarray}
K_{\rm M}(\tilde Z) = u_\ast \, L_{\rm O}  \, {\rm Ri}_{\rm f}(\tilde Z) ,
\label{DG4}
\end{eqnarray}
where $K_{\rm M} = \kappa_0 \, u_\ast \, z$ for $ |\tilde Z|  \ll 1$ and $K_{\rm M} = u_\ast \, |L_{\rm O}|  \, \tilde Z^{4/3}$ for $ |\tilde Z|  \gg 1$;
}
\item{
the turbulent Prandtl number,
\begin{eqnarray}
{\rm Pr}_{_{\rm T}}(\tilde Z) = {\rm Pr}_{_{\rm T}}^{(0)}  \left[1 - { C_\theta \, C_{\rm p} \,{\rm Ri}_{\rm f}(\tilde Z)
\over A_z \left(1 - {\rm Ri}_{\rm f}(\tilde Z) \right)}
\right]^{-1} ,
\nonumber\\
\label{DG6}
\end{eqnarray}
where ${\rm Pr}_{_{\rm T}} = {\rm Pr}_{_{\rm T}}^{(0)}$ for $ |\tilde Z|  \ll 1$ and ${\rm Pr}_{_{\rm T}} = {\rm Pr}_{_{\rm T}}^{(\infty)} \equiv{\rm Pr}_{_{\rm T}}^{(0)}/(1+C_\theta \, C_{\rm p})$ for $ |\tilde Z|  \gg 1$;
}
\item{
 the level of temperature fluctuations,
\begin{eqnarray}
E_\theta(\tilde Z) = \theta_\ast^2 \,  C_{\rm p} \, (2 C_\tau \, A_z)^{-1/2} \,   \, {{\rm Pr}_{_{\rm T}}(\tilde Z) \over
 \tilde  E_{\rm K}(\tilde Z)}  ,
 \label{DG7}
\end{eqnarray}
where $E_\theta = \theta_\ast^2 \, C_{\rm p} \, \left(2 C_\tau \, A_z^{(0)}\right)^{-1/2}
\,   {\rm Pr}_{_{\rm T}}^{(0)}$ for $ |\tilde Z|  \ll 1$ and $E_\theta = \theta_\ast^2 \, C_{\rm p} \, \, (2 C_\tau)^{-1/2}
\,   {\rm Pr}_{_{\rm T}}^{(\infty)} \,  \tilde Z^{-2/3}$ for $ |\tilde Z|  \gg 1$;
}
\item{
the large-scale shear,
\begin{eqnarray}
\tilde S(\tilde Z) = {u_\ast \over L_{\rm O} \, {\rm Ri}_{\rm f}(\tilde Z)}  ,
\label{DG3}
\end{eqnarray}
where $\tilde S(\tilde Z)   =u_\ast  /(\kappa_0 \, z)$ for $ |\tilde Z|  \ll 1$ and $\tilde S(\tilde Z)   = (u_\ast  / |L_{\rm O}|)\, \tilde Z^{-4/3}$ for $ |\tilde Z|  \gg 1$;
}
\item{
the vertical gradient of the mean potential temperature,
\begin{eqnarray}
\nabla_z \overline{\Theta}^{({\rm m})} ={ \theta_\ast \,  {\rm Pr}_{_{\rm T}}(\tilde Z)  \over |L_{\rm O}|  \, {\rm Ri}_{\rm f}(\tilde Z) } ,
 \label{DG8}
\end{eqnarray}
where $\nabla_z \overline{\Theta}^{({\rm m})}  =-  \theta_\ast  \, {\rm Pr}_{_{\rm T}}^{(0)} / (\kappa_0 \, z)$ for $ |\tilde Z|  \ll 1$ and $\nabla_z \overline{\Theta}^{({\rm m})}  = - (\theta_\ast / |L_{\rm O}|) \, {\rm Pr}_{_{\rm T}}^{(\infty)} \,  \tilde Z^{-4/3}$ for $ |\tilde Z|  \gg 1$.
}
\end{itemize}
Here ${\rm Ri}_{\rm f} = - \beta \, \tilde F_z / (K_{\rm M} \tilde S^2)$ is the flux Richardson number,
$\tilde F_z$ is the vertical turbulent flux of the potential temperature at the surface layer,
$u_\ast^2=K_{\rm M} \, \tilde S$ with $u_\ast$ being the local ($z$-dependent) friction velocity,
$\tilde S = d\meanU_{\rm h}/dz$ is the large-scale shear  at the surface layer, $\meanU_{\rm h}(z)$ is the mean
horizontal velocity at the surface layer,
$\theta_\ast = |\tilde F_z|/u_\ast = u_\ast^2 / \beta \, |L_{\rm O}|$ with $L_{\rm O}$ being the local Obukhov length
defined as $L_{\rm O}= - u_\ast^{3} / (\beta \, \tilde F_z)$,
the normalized height $\tilde Z = \kappa_0 \, z / L_{\rm O}$ with
$\kappa_0=0.4$ being the von Karman constant, and ${\rm Pr}_{_{\rm T}}^{(0)} = C_\tau / C_{\rm F}$ is the
turbulent Prandtl number for a non-stratified turbulence at ${\rm Ri}_{\rm f}=0$.
Note that the flux Richardson number ${\rm Ri}_{\rm f}$ is negative in the convective turbulence
in surface layer, and its absolute value is not limited and can be large.
The local Obukhov length $L_{\rm O}$ is negative in the convective turbulence as well,
but the product $L_{\rm O} \, {\rm Ri}_{\rm f}$ is positive.
The vertical profile of the normalized turbulent kinetic energy density,
$\tilde E_{\rm K}(\tilde Z) = E_{\rm K}(\tilde Z)/E_{\rm K0}$, is determined
by the following  nonlinear equation:
\begin{eqnarray}
\tilde E_{\rm K}^2 + \tilde Z \, \tilde E_{\rm K}^{1/2} - 1 =0 ,
 \label{DG11}
\end{eqnarray}
where $E_{\rm K0} = u_\ast^2 / (2 C_\tau \, A_z)^{1/2}$,
and $\tilde E_{\rm K} = 1 +  |\tilde Z| / 2$ for $ |\tilde Z|  \ll 1$ and
$\tilde E_{\rm K} = \tilde Z^{2/3}$ for $ |\tilde Z|  \gg 1$.
Using Eq.~(\ref{DG3}), we obtain the mean horizontal velocity at the surface layer as
\begin{eqnarray}
\meanU_{\rm h}(z) = {u_\ast \over \kappa_0} \int_0^z {d \tilde Z \over \tilde Z \,
\left[\tilde E_{\rm K}(\tilde Z)\right]^{1/2}} .
\label{FV0}
\end{eqnarray}
For illustration, the vertical profile~(\ref{FV0}) of the normalized mean horizontal velocity
 $\meanU_{\rm h}(z)/u_\ast$ at the surface layer is shown in  Fig.~\ref{Fig11},
 where we use the numerical solution of nonlinear equation~(\ref{DG11}) (see Ref.~\cite{RKZ22}).

 \subsection{Matching of solutions for CBL core and convective surface layer}
\label{subsec:4.2}

The matching between the solutions obtained for the CBL core and the convective surface layer
are performed as follows.
\begin{itemize}
\item{
The mean potential temperature $\overline{\Theta}^{({\rm m})}$ and the mean horizontal velocity $\meanU_{\rm h}$
for the convective surface layer  at $|\tilde Z| \gg 1$ are matched with the mean potential temperature
$\overline{\Theta}^{({\rm s})}$ and the mean velocity $\overline{U}_r^{({\rm s})}$
for the CBL core at $\pi z / L_z \ll 1$.
}
\item{
The vertical turbulent flux of potential temperature $\tilde F_z$
for the convective surface layer at $|\tilde Z| \gg 1$ are matched with
the vertical flux of potential temperature,
$\left\langle \overline{\Theta}^{({\rm s})} \meanU_{z}^{({\rm s})} \right\rangle_{_{V}} + \langle F_z \rangle_{_{V}}$,
transported by both, the semi-organized structures and turbulence at $\pi z / L_z \ll 1$.
}
\end{itemize}

\begin{figure}
\vspace*{1mm}
\centering
\includegraphics[width=9.5cm]{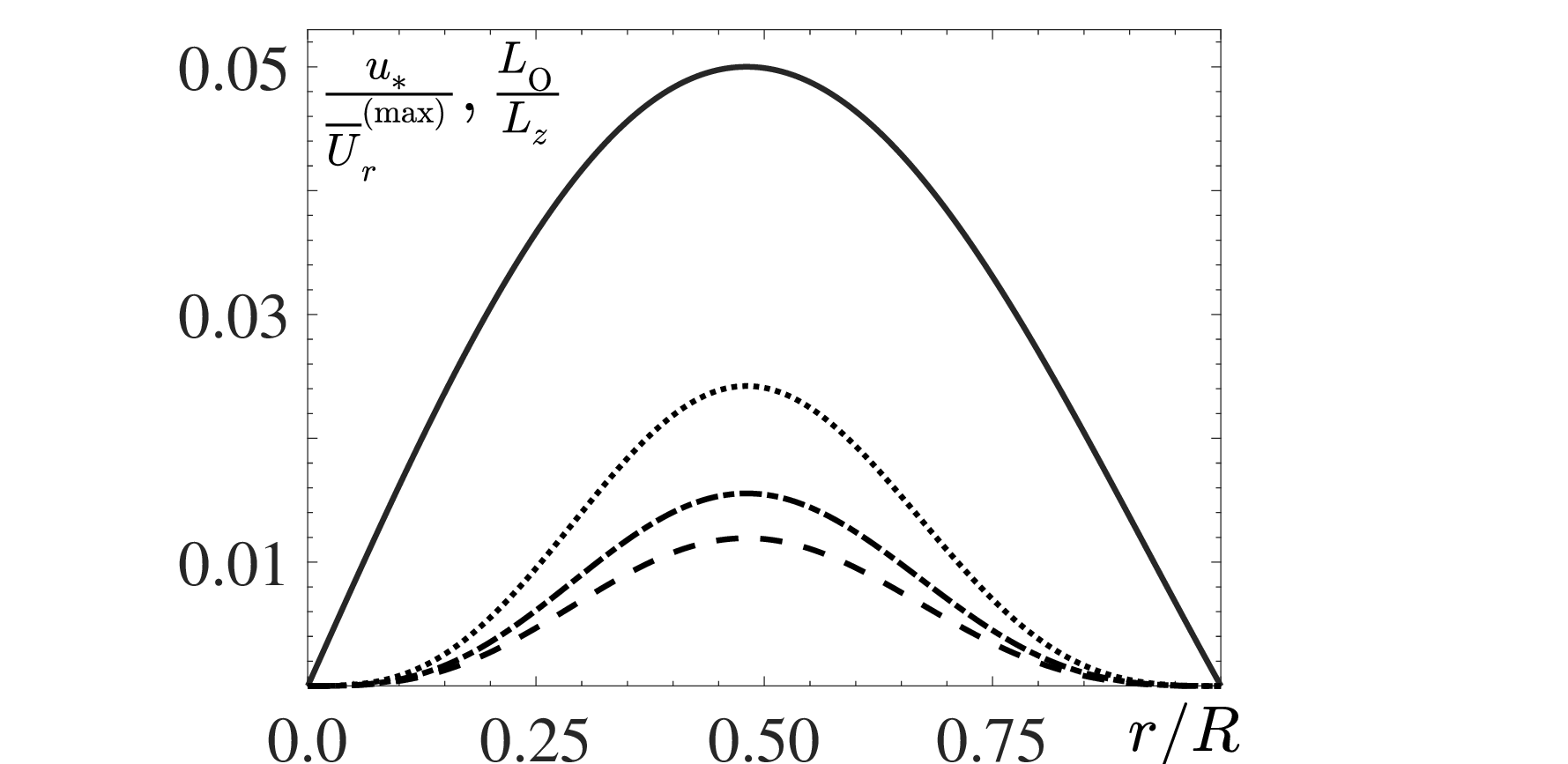}
\caption{\label{Fig12} The radial profiles of the ratio $L_{\rm O}(r)/L_z$
for different values of the scale separation parameter $L_z/\ell_z=$
$7$  (dashed line);  $8$ (dashed-dotted line) and $10$ (dotted line);
the radial profile of the ratio $u_\ast(r) /\meanU_r^{\rm \, (max)}$ (solid line).
}
\end{figure}

As the result, the ratio of the local Obukhov length scale, $L_{\rm O}= - u_\ast^{3} / (\beta \, \tilde F_z)$,
for the convective surface layer to the vertical size of the semi-organized structure $L_z$
is $L_{\rm O}/L_z = - (u_\ast/U_{\rm D})^{3}$, where
the convective velocity $U_{\rm D}$ in this case is $U_{\rm D}=\left(\beta \, \tilde F_z \, L_z\right)^{1/3}$,
and we take into account that
$\left|\left\langle \overline{\Theta}^{({\rm s})} \meanU_{z}^{({\rm s})}
\right\rangle_{_{V}}\right| \gg \left|\langle F_z \rangle_{_{V}}\right|$.
Using Eqs.~(\ref{BB6}), (\ref{TTP14}) and~(\ref{FV0}), we obtain
the ratio of the friction velocity $u_\ast$ to the convective velocity $U_{\rm D}$ as
\begin{eqnarray}
{u_\ast \over U_{\rm D}} = {\kappa_0 \, A_\ast \, \left(1 - \langle \hat F \rangle_{_{V}}\right)^{1/6}  \over C_\tau^{1/2} \, f_{F_{s}}^{1/3}(A_\ast) \,  I_{\rm U} } \, \left({L_z \over \ell_z}\right)^{2\over 3} \,
\,  J_1\left({\lambda \, r \over R}\right) ,
\nonumber\\
\label{FV1}
\end{eqnarray}
where $\meanU_{\rm h}\left(|\tilde Z| \gg 1\right) \approx u_\ast \,  I_{\rm U}/ \kappa_0$ and
\begin{eqnarray}
I_{\rm U} =\int_0^\infty {d \tilde Z \over \tilde Z \, \left[\tilde E_{\rm K}(\tilde Z)\right]^{1/2}} .
\label{FV3}
\end{eqnarray}
The ratio of the friction velocity $u_\ast$ to the maximum value $\meanU_r^{\rm \, (max)}$ of
the horizontal velocity of the semi-organized structure $\meanU_{r}^{({\rm s})}$
is given by
\begin{eqnarray}
{u_\ast(r) \over \meanU_r^{\rm \, (max)}} = { \kappa_0 \over I_{\rm U} } \, \, {J_1(\lambda \, r/ R) \over J_1(\lambda \, r_{\rm max}/ R)} ,
\label{FV2}
\end{eqnarray}
where $\meanU_r^{\rm \, (max)}=\meanU_{r}^{({\rm s})}\left(r=r_{\rm max}, z \to 0\right)$ and $r_{\rm max}$ is the radius at which the function $J_1(Y)$ reaches the maximum value.
In Fig.~\ref{Fig12} we show the radial profiles of the ratios $L_{\rm O}/L_z$ and $u_\ast /\meanU_r^{\rm \, (max)}$,
which demonstrate that these ratios are small.
Therefore, this analysis allows us to connect the global turbulent characteristics in the atmospheric CBL-core
with the basic characteristics  of the convective surface layer.

\section{Conclusions}
\label{sec:5}

In the present study we investigate essential features of turbulence and semi-organized structures
in the core of the atmospheric convective boundary-layer (the CBL core)
by means of the energy- and flux budget (EFB) theory.
\begin{itemize}
\item{
Using the analytical solution~(\ref{BB6})--(\ref{BB10}) for the semi-organized structures
and budget equations~(\ref{C1})--(\ref{C15}) for the basic second moments in convective turbulence,
we find the global turbulent characteristics
(averaged over the entire volume of the semi-organized structure) including turbulent kinetic energy density
and intensity of potential temperature fluctuations, turbulent flux of potential temperature, as well as
the kinetic and thermal energy densities of the semi-organized structures.
}
\item{
Applying the analytical description for plumes based on the two-point instantaneous correlation functions
$\langle \theta(t,{\bm x}) \, u_z(t,{\bm x}+{\bm r}) \rangle$,
we connect the global characteristics of convective turbulence and semi-organized structures
with degree of the thermal anisotropy $\alpha$ defined by Eq.~(\ref{TD6}) and characterized the form of plumes.
}
\item{
We demonstrate that when plumes are extended in the vertical direction
($\ell_{\rm h}^{\rm (pl)} /\ell_{z}^{\rm (pl)} < 1$),  the vertical turbulent flux of
potential temperature averaged over the entire volume of the semi-organized structure $\langle F_z \rangle_{_{V}}$ is negative
(see Fig.~\ref{Fig3}), where $\ell_{\rm h}^{\rm (pl)} $ and $\ell_{z}^{\rm (pl)} $ are the horizontal and vertical
scales in which the two-point instantaneous correlation functions $\langle \theta(t,{\bm x}) \, u_z(t,{\bm x}+{\bm r}) \rangle $
characterized the plumes, vanish in the horizontal and vertical directions, respectively.
}
\item{
When $\ell_{\rm h}^{\rm (pl)} /\ell_{z}^{\rm (pl)} < 1$, the  mean vertical gradient of the potential temperature $\nabla_z \overline{\Theta}^{({\rm m})}$ is positive inside the semi-organized structure [see Eq.~(\ref{TP31}) and Fig.~\ref{Fig10}].
The  gradient $\nabla_z \overline{\Theta}^{({\rm m})}$ increases with
decrease of the scale-separation parameter $L_z/\ell_z$.
}
\item{
We demonstrate that the vertical flux of potential temperature
$\left\langle \overline{\Theta}^{({\rm s})} \meanU_{z}^{({\rm s})} \right\rangle_{_{V}}$ transported by the semi-organized structures is much larger than the vertical turbulent flux $\langle F_z \rangle_{_{V}}$  of potential temperature
[see Eq.~(\ref{TP14}) and Fig.~\ref{Fig8}].
}
\item{
The turbulent kinetic energy density $\langle E_{\rm K} \rangle_{_{V}}$
is much smaller than the kinetic energy density of semi-organized structures
$\langle  E_{\rm U} \rangle_{_{V}}$ [see Eq.~(\ref{TP28}) and Fig.~\ref{Fig7}].
Increase of the scale separation between the vertical size $L_z$ of the semi-organized
structures and the vertical integral scale of turbulence $\ell_z$,
increases the kinetic energy density $\langle  E_{\rm U} \rangle_{_{V}}$ of the semi-organized structures
and the ratio $\langle  E_{\rm U} \rangle_{_{V}}/\langle E_{\rm K} \rangle_{_{V}}$.
}
\item{
The turbulent thermal energy density $\langle E_\theta \rangle_{_{V}}$ is much smaller than
the thermal energy  density of the semi-organized structures $\langle E_\Theta \rangle_{_{V}}$
[see Eq.~(\ref{TP29}) and Fig.~\ref{Fig9}].
}
\item{
The global turbulence characteristics depend on the aspect ratio of the semi-organized structure,
the scale separation parameter between the vertical size $L_z$ of the structures and the integral scale of turbulence $\ell_z$,
and the degree of thermal anisotropy (i.e., the form of plumes).
In the limit of large aspect ratio of the semi-organized structures,
these global turbulence characteristics reaches their asymptotic values which depend
on the degree of thermal anisotropy and the scale separation parameter.
}
\item{
We connect the global turbulent characteristics in the atmospheric CBL-core
with the basic characteristics  of the convective surface layer.
This analysis yields the ratio of the local Obukhov length scale, $L_{\rm O}= - u_\ast^{3} / (\beta \, \tilde F_z)$,
for the convective surface layer to the vertical size of the semi-organized structure $L_z$
as well as the ratio of the friction velocity $u_\ast$ to the maximum value of
the horizontal velocity of the semi-organized structure $\meanU_{r}^{({\rm s})}$
[see Eqs.~(\ref{FV1}) and~(\ref{FV2}), as well as Fig.~\ref{Fig12}].
}
\end{itemize}
The obtained theoretical results are important for modeling applications
in the atmospheric convective boundary-layer.
These results are also very useful for analysis of laboratory and field experiments,
direct numerical simulations as well as large-eddy simulations of convective turbulence
with large-scale semi-organized structures.

\bigskip
\bigskip

\noindent
{\bf DEDICATION}

This paper was dedicated to Prof. Sergej Zilitinkevich (1936-2021) who initiated this work and discussed some of the obtained results.

\bigskip
\medskip
\noindent
{\bf ACKNOWLEDGMENTS}

This research was supported in part by the PAZY Foundation of the Israel Atomic Energy Commission (Grant No. 122-2020).

\medskip
\noindent
{\bf AUTHOR DECLARATIONS}

\noindent
{\bf Conflict of Interest}

The authors have no conflicts to disclose.

\medskip
\noindent
{\bf DATA AVAILABILITY}
\medskip

Data sharing is not applicable to this article as no new data were
created or analyzed in this study.

\bigskip
\noindent
{\bf NOMENCLATURE}
\medskip
\\
\noindent
$A_{x,y}=E_{x,y}/E_{\rm K}$ horizontal  anisotropy parameters
\\
\noindent
$A_z=E_{z}/E_{\rm K}$ vertical anisotropy parameter
\\
\noindent
$A_\ast=\pi \, R / (\lambda \,  L_z)$ aspect ratio
of the semi-organized structures
\\
\noindent
$E_{\rm K}=\langle {\bm u}^2 \rangle/2$  density of turbulent kinetic energy (TKE)
\\
\noindent
$\tilde E_{\rm K} = E_{\rm K}/E_{\rm K0}$ normalized density of turbulent kinetic energy
\\
\noindent
$E_{\rm K0} = u_\ast^2 / (2 C_\tau \, A_z)^{1/2}$ density of turbulent kinetic energy at the surface
\\
\noindent
$E_z=\langle u_z^2 \rangle/2$ density of the vertical turbulent kinetic energy
\\
\noindent
$E_\alpha = \langle u_\alpha^2\rangle/2$ horizontal and vertical turbulent kinetic energies ($\alpha=x,y,z$)
\\
\noindent
$\langle  E_{\rm U} \rangle_{_{V}}$ kinetic energy density of the semi-organized structure
\\
\noindent
$E_\theta=\langle \theta^2 \rangle/2$ intensity of potential temperature fluctuations
\\
\noindent
$\langle E_\Theta \rangle_{_{V}}$ thermal energy density of semi-organized structure
\\
\noindent
$F_i = \langle u_i \, \theta \rangle$  turbulent flux of potential temperature
\\
\noindent
$F_{x,y}$ horizontal turbulent flux of potential temperature
\\
\noindent
$F_z=\langle u_z \, \theta \rangle$ vertical turbulent flux of potential temperature
\\
\noindent
$\langle F_z^{({\rm D})} \rangle_{_{V}}$ Deerdorf turbulent flux
\\
\noindent
$\langle F_z^{({\rm W})} \rangle_{_{V}} $ secondary vertical turbulent flux
\\
\noindent
$\langle \hat F \rangle_{_{V}}\equiv \beta \, \langle F_z \rangle_{_{V}} \, \ell / \langle E_{\rm K}^{3/2} \rangle_{_{V}}$
normalized  averaged vertical turbulent flux of the potential temperature
\\
\noindent
${\bm g}$  gravity acceleration
\\
\noindent
$J_m(x)$ Bessel function of the first kind
\\
\noindent
$K_{\rm H}$ turbulent heat conductivity
\\
\noindent
$K_{\rm M}$ turbulent (eddy) viscosity
\\
\noindent
$L_{\rm O}= - \tau^{3/2} / (\beta \, F_z)$ local Obukhov length
\\
\noindent
$L_z$ height of the semi-organized structure
\\
\noindent
$\ell_z$ vertical integral scale
\\
\noindent
$\ell_{\rm h}^{\rm (pl)}$ characteristic horizontal scale of plumes
\\
\noindent
$\ell_{z}^{\rm (pl)}$ characteristic vertical scale of plumes
\\
\noindent
$N = (\beta \, |\nabla_z \overline{\Theta}|)^{1/2}$  Brunt-V\"{a}is\"{a}l\"{a} frequency
\\
\noindent
$Q_{ij} = \rho_0^{-1} (\langle p \nabla_i u_j\rangle + \langle p \nabla_j u_i\rangle)$ inter-component
energy exchange term
\\
\noindent
$Q_{\alpha\alpha} = 2 \rho_0^{-1} (\langle p \nabla_\alpha u_\alpha\rangle$ diagonal terms of the tensor $Q_{ij}$
\\
 \noindent
 $p$ fluctuations of the fluid pressure
\\
 \noindent
 $P=\overline{P}^{({\rm m})} + \overline{P}^{({\rm s})} + p$ total fluid pressure with the reference value $P_\ast$
\\
 \noindent
 $\overline{P}^{({\rm m})}$  mean fluid pressure
\\
 \noindent
 $\overline{P}^{({\rm s})}$ mean fluid pressure related to the large-scale semi-organized structures
\\
\noindent
${\rm Pr}_{_{\rm T}} = K_{\rm M} / K_{\rm H}$ turbulent Prandtl number
\\
\noindent
${\rm Pr}_{_{\rm T}}^{(0)} = C_\tau / C_{\rm F}$ turbulent Prandtl number for a non-stratified turbulence
\\
\noindent
$R$ radius of the semi-organized structure
\\
\noindent
${\rm Ri} = N^2 / S^2$ gradient Richardson number
\\
\noindent
${\rm Ri}_{\rm f} = - \beta \, F_z / (K_{\rm M} S^2)$ flux Richardson number
\\
\noindent
$R_\infty={\rm Ri}_{\rm f}({\rm Ri} \to \infty)$ flux Richardson number at very large gradient Richardson number
\\
\noindent
$S$ mean velocity shear caused by semi-organized structure
\\
\noindent
$\tilde S$ mean velocity shear in the surface layer
\\
\noindent
$t_{\rm T}= \ell_z /E_{z}^{1/2}$ turbulent dissipation timescale
\\
\noindent
$T$ fluid temperature with the reference value $T_\ast$
\\
\noindent
${\bm u}=(u_x, u_y, u_z)$ fluctuations of the fluid velocity
\\
\noindent
$u_\ast$ local ($z$-dependent) friction velocity
\\
 \noindent
 ${\bm U}=\overline{\bm U}^{({\rm m})} + \overline{\bm U}^{({\rm s})} + {\bm u}$ total velocity
\\
  \noindent
 $U_{\rm D}$  characteristic convective velocity
\\
 \noindent
 $\meanUU= \overline{\bm U}^{({\rm m})} + \overline{\bm U}^{({\rm s})}$ total mean velocity
\\
 \noindent
 $\meanUU^{({\rm s})}$ mean velocity related to the semi-organized structure
\\
 \noindent
 $\meanUU^{({\rm m})}(z)=(\meanU_x, \meanU_y, 0)$ mean-wind velocity
\\
\noindent
$\meanUU_{\rm h}^{({\rm s})}$ horizontal component of mean velocity related to the semi-organized structure
\\
\noindent
$\meanUU_{z}^{({\rm s})}$ vertical component of mean velocity related to the semi-organized structures
\\
  \noindent
 $\meanWW^{({\rm s})} =\bec{\nabla} {\bm \times}  \meanUU^{({\rm s})}$ mean vorticity characterized the semi-organized structure
\\
\noindent
$\tilde Z = \kappa_0 \, z / L_{\rm O}$ normalized height
\\
\noindent
$\alpha$ degree of thermal anisotropy
\\
\noindent
$\beta=g/T_\ast$ buoyancy parameter
\\
\noindent
$\gamma = c_{\rm p}/c_{\rm v}$ specific heat ratio
\\
\noindent
$\delta_{ij}$ Kronecker unit tensor
\\
\noindent
$\varepsilon_i^{({\rm F})} = (\nu + \chi) \, \langle (\nabla_j u_i)
\, (\nabla_j \theta) \rangle$  dissipation rate of the turbulent heat flux
\\
\noindent
$\varepsilon_{_{\rm K}} = \nu \, \langle (\nabla_j u_i)^2 \rangle$ dissipation rate of $E_{\rm K}$
\\
\noindent
$\varepsilon_{\alpha} = \nu \, \langle (\nabla_j u_\alpha)^2 \, \rangle$ dissipation rate of horizontal and vertical turbulent kinetic energy density components $E_\alpha$
\\
\noindent
$\varepsilon_\theta = \chi \, \langle ({\bm \nabla} \theta)^2 \rangle$
dissipation rate of $E_\theta$
\\
\noindent
$\varepsilon_{i}^{(\tau)} = \varepsilon_{iz}^{(\tau)} - \beta \, F_i - Q_{iz}$
effective dissipation rate of the off-diagonal components of the Reynolds stress
\\
\noindent
$\varepsilon_{iz}^{(\tau)}=2 \nu \, \langle (\nabla_j u_i) \, (\nabla_j u_z) \rangle$
molecular-viscosity dissipation rate of the off-diagonal components of the Reynolds stress
\\
\noindent
$\theta$  fluctuations of the potential temperature
\\
\noindent
$\theta_\ast = |F_z|/u_\ast = u_\ast^2 / (\beta \, |L_{\rm O}|)$ level of potential temperature fluctuations
\\
 \noindent
 $\Theta = T (P_\ast / P)^{1-\gamma^{-1}} =\overline{\Theta}^{({\rm m})} + \overline{\Theta}^{({\rm s})} + \theta$ total  potential temperature
\\
 \noindent
$\Theta_{\rm D}$  characteristic convective temperature
\\
 \noindent
 $\overline{\Theta}^{({\rm m})}(z)$ mean potential temperature
\\
 \noindent
 $\overline{\Theta}^{({\rm s})}$ mean potential temperature related to the large-scale semi-organized structures
\\
 \noindent
$\tilde \Theta = \sqrt{2 \langle  E_\Theta \rangle_{_{V}}}/\Theta_{\rm D}$  normalized potential temperature
\\
\noindent
$\kappa_0=0.4$ von Karman constant
\\
\noindent
$\lambda=3.83$  first root of equation $J_1(x)=0$
\\
\noindent
$\mu=(2 \alpha+3)/(8 \alpha-3)$ parameter characterizing the form of semi-organized structures
\\
\noindent
$\nu$  kinematic viscosity of fluid
\\
\noindent
$\Pi_{\rm K} = - \tau_{i z} \, \nabla_z \meanU_i = K_{\rm M} \, S^2$ production rate of the turbulent kinetic energy density
\\
\noindent
$\rho_0$  fluid density
\\
\noindent
$\sigma= 4 \, (8 \alpha-3)/45$ parameter characterizing the form of semi-organized structures
\\
\noindent
$\tau_{iz} =\langle u_i \, u_z \rangle$  off-diagonal components of the Reynolds
stress and $i=x, y$
\\
\noindent
$\tau = \left(\tau_{xz}^2 + \tau_{yz}^2\right)^{1/2} = K_{\rm M} \, S  \equiv u_\ast^2$ vertical turbulent flux of
momentum
\\
\noindent
${\bm \Phi}_{i}^{({\rm F})} = \langle u_i \, u_z \, \theta\rangle - \nu \, \langle \theta \, (\nabla_z u_i) \rangle
- \chi\, \langle u_i \, (\nabla_z \theta) \rangle$ flux of $F_i$
\\
\noindent
$\Phi_{i}^{(\tau)}=\langle u_i \, u_z^2 \rangle + \rho_0^{-1} \, \langle p \, u_i\rangle - \nu \, \nabla_z \tau_{iz}$ flux of $\tau_{iz}$
\\
\noindent
$\Phi_{\rm K} = \rho_0^{-1} \langle u_z \, p\rangle + (\langle u_z
\, {\bf u}^2 \rangle - \nu \, \nabla_z \langle {\bf u}^2 \rangle)/2$ flux of $E_{\rm K}$
\\
\noindent
$\Phi_{x,y}= (\langle u_z \, u_{x,y}^2 \rangle - \nu \, \nabla_z \langle u_{x,y}^2 \rangle) / 2$ flux of
the horizontal turbulent kinetic energy components $E_{x,y}$
\\
\noindent
$\Phi_{z}= \rho_0^{-1} \langle u_z \, p\rangle + (\langle u_z^3 \rangle - \nu \, \nabla_z \langle u_z^2 \rangle) / 2$
flux of the vertical turbulent kinetic energy components $E_{z}$
\\
\noindent
$\Phi_{\alpha}$ flux of $E_\alpha$ with $\alpha=x,y,z$
\\
\noindent
$\Phi_\theta = \left(\langle u_z \, \theta^2 \rangle - \chi \, \nabla_z \langle  \theta^2 \rangle\right)/2$
flux of $E_\theta$
\\
\noindent
$\chi$ molecular temperature diffusivity
\\
\noindent
$\Psi$  stream-function

\appendix

\section{Functions $f(A_\ast)$}
\label{Appendix:A}

\begin{eqnarray}
f_{\hat F}(A_\ast) = {(2 \pi)^2 \over 3} \, {\left(1 + A_\ast^2\right)^2 \over A_\ast^2 \, \sigma \, \left(A_\ast^2-\mu\right)} ,
\label{AP2}
\end{eqnarray}

\begin{eqnarray}
\tilde f_S(A_\ast) &=&\lambda^2 \, J_0^2(\lambda) \, \left(1 +  A_\ast^2\right)^2
+ 2 A_\ast^2 \left[J_0^2(\lambda) -1\right],
\nonumber\\
\label{AP1a}
\end{eqnarray}

\begin{eqnarray}
f_S(A_\ast) &=&\left({\pi \over \lambda \, A_\ast} \right)^2 \, \tilde f_S(A_\ast) ,
\label{AP1}
\end{eqnarray}

\begin{eqnarray}
f_{F_{s}}(A_\ast) ={\pi^2 \over \sqrt{3}}  \, J_0^2(\lambda) \, {\left(1 + A_\ast^2\right)^2 \over A_\ast^2} \, f_S^{1/2}(A_\ast)  ,
\label{AP4}
\end{eqnarray}

\begin{eqnarray}
f_U(A_\ast) = {J_0^2(\lambda) \over  4} \, {(1 + A_\ast^2) \over f_{F_{s}}^{2/3}(A_\ast)} ,
\label{AP5}
\end{eqnarray}

\begin{eqnarray}
f_\Theta(A_\ast) = {f_{F_{s}}^{2/3}(A_\ast) \over J_0^2(\lambda)} ,
\label{AP6}
\end{eqnarray}

\begin{eqnarray}
f_u(A_\ast) ={f_S(A_\ast) \over  f_{F_{s}}^{2/3}(A_\ast) } ,
\label{AP7}
\end{eqnarray}

\begin{eqnarray}
f_{F_{T}}(A_\ast) ={1 \over \sqrt{3}} \, {f_{\hat F}(A_\ast) \, f_S^{3/2}(A_\ast) \over  f_{F_{s}}(A_\ast)},
\label{AP3}
\end{eqnarray}

\begin{eqnarray}
f_\theta(A_\ast) =8 \pi^2 \,  {1 + A_\ast^2 \over A_\ast^2} \, f_\Theta(A_\ast) ,
\label{AP8}
\end{eqnarray}

\begin{eqnarray}
f_D(A_\ast) =2\, \sqrt{3} \, {f_\theta(A_\ast) \over f_{F_{T}}(A_\ast) \, f_u^{1/2}(A_\ast) } ,
\label{AP9}
\end{eqnarray}

\begin{eqnarray}
f_{_{\nabla}}(A_\ast) ={f_{F_{T}}(A_\ast) \over 2\, \sqrt{3} \, f_u^{1/2}(A_\ast) } .
\label{AP10}
\end{eqnarray}

\section{Identities used for derivation of  Eqs.~(\ref{TP9}), (\ref{TTP12}), (\ref{TP23}), (\ref{TP27}), (\ref{TP21}) and~(\ref{TP31})}
\label{Appendix:B}

To determine the volume averaged production rate of the turbulent kinetic energy, we use the analytical
solution~(\ref{BB6})--(\ref{BB7}) for the velocity $\meanUU^{({\rm s})}$ of the semi-organized structure.
In particular, the squared large-scale shear $\left\langle S^2\right\rangle_{_{V}}$
averaged over the entire volume of the semi-organized structure
is given by
\begin{eqnarray}
&& \left\langle S^2\right\rangle_{_{V}} = \left\langle \left(\nabla_r \meanU_r^{({\rm s})}\right)^2\right\rangle_{_{V}} + \left\langle\left(\nabla_z \meanU_z^{({\rm s})}\right)^2\right\rangle_{_{V}}
\nonumber\\
&& \;+ {1 \over 2}  \left\langle\biggl(\nabla_r \meanU_z^{({\rm s})} + \nabla_z \meanU_r^{({\rm s})} \biggr)^2\right\rangle_{_{V}}
=  {1 \over 4}\, \left({\meanU_{z0} \over R}\right)^2 \, \tilde f_S(A_\ast) ,
\nonumber\\
\label{TP1}
\end{eqnarray}
where the function $\tilde f_S(A_\ast)$ is given by Eq.~(\ref{AP1a}) in Appendix~\ref{Appendix:A}.
To derive Eq.~(\ref{TP1}), we use the following equations:
\begin{eqnarray}
&& \left\langle \left(\nabla_r \meanU_{r}^{({\rm s})}\right)^2 \right\rangle_{_{V}} =
{1 \over 2}\, \left({\meanU_{z0} \over R}\right)^2 \, A_\ast^2 \, \left[(1 + \lambda^2)\, J_0^2(\lambda) -1\right] ,
\nonumber\\
\label{TTPP1}
\end{eqnarray}
\begin{eqnarray}
&& \left\langle \left(\nabla_z \meanU_{z}^{({\rm s})}\right)^2 \right\rangle_{_{V}} =
- \left\langle \left(\nabla_r \meanU_{z}^{({\rm s})}\right) \,  \left(\nabla_z \meanU_{r}^{({\rm s})}\right) \right\rangle_{_{V}}
\nonumber\\
&& = {1 \over 2}\, \left({\meanU_{z0} \over R}\right)^2 \, A_\ast^2 \, \lambda^2 \, J_0^2(\lambda) ,
\label{TTPP2}
\end{eqnarray}
\begin{eqnarray}
&& \left\langle \left(\nabla_z \meanU_{r}^{({\rm s})}\right)^2 \right\rangle_{_{V}} =
{1 \over 2}\, \left({\meanU_{z0} \over R}\right)^2 \, A_\ast^4 \, \lambda^2 \, J_0^2(\lambda) ,
\label{TTPP3}
\end{eqnarray}
\begin{eqnarray}
&& \left\langle \left(\nabla_r \meanU_{z}^{({\rm s})}\right)^2 \right\rangle_{_{V}} =
{1 \over 2}\, \left({\meanU_{z0} \over R}\right)^2 \,\lambda^2 \, J_0^2(\lambda) ,
\label{TTPP4}
\end{eqnarray}
where we use integrals given by Eqs.~(\ref{BP1})--(\ref{BP4}).
Equation~(\ref{TP1}) allows us to derive Eq.~(\ref{TP9}) for the turbulent kinetic energy density $\langle E_{\rm K} \rangle_{_{V}}$.

To derive Eq.~(\ref{TTP12}) for the vertical flux of potential temperature, $\left\langle \overline{\Theta}^{({\rm s})} \meanU_{z}^{({\rm s})} \right\rangle_{_{V}}$, transported by the semi-organized structures, we use the identity:
\begin{eqnarray}
&& \left\langle \overline{\Theta}^{({\rm s})} \meanU_{z}^{({\rm s})} \right\rangle_{_{V}} = {1 \over 2}
 \overline{\Theta}_{0} \, \meanU_{z0} \, J_0^2(\lambda)  ,
\label{TTTP12}
\end{eqnarray}
and apply Eqs.~(\ref{BB7}), (\ref{BB8}), and~(\ref{BP1}).

To derive Eq.~(\ref{TP23}) for the kinetic energy
density of semi-organized structures, we use the following identity:
\begin{eqnarray}
\left\langle \left( \meanU_{z}^{({\rm s})} \right)^2  \right\rangle_{_{V}} &=&
A_\ast^{-2} \, \left\langle \left( \meanU_{r}^{({\rm s})} \right)^2  \right\rangle_{_{V}}
\nonumber\\
&=& { \meanU_{z0}^2 \over 2} \, J_0^2(\lambda) ,
\label{TTTP1}
\end{eqnarray}
where we apply Eqs.~(\ref{BB6}), (\ref{BB7}), and~(\ref{BP1}).
To derive Eq.~(\ref{TP27}) for the thermal energy density of the semi-organized structures, we use the identity:
\begin{eqnarray}
&& \left\langle \left(\overline{\Theta}^{({\rm s})}\right)^2 \right\rangle_{_{V}}
= \overline{\Theta}_{0}^2 \, {J_0^2 (\lambda) \over 2 } ,
\label{BP6}
\end{eqnarray}
where we  apply Eqs.~(\ref{BB8}) and~(\ref{BP1}).

Equation~(\ref{TP24}) for the turbulent kinetic energy density $\langle E_{\rm K} \rangle_{_{V}}$ allows us to determine
the turbulent time $t_{\rm T}=\ell / \sqrt{\langle E_{\rm K} \rangle_{_{V}}}$,
\begin{eqnarray}
t_{\rm T} &=& \sqrt{3} \,  {L_z \over U_{\rm D}} \, \left({\ell_z \over L_z}\right)^{2\over 3}
\, { \left(1 - \langle \hat F \rangle_{_{V}}\right)^{1/3} \over f_u^{1/2}(A_\ast)  } ,
\label{TP26}
\end{eqnarray}
which is much smaller than the characteristic convective time $L_z /U_{\rm D}$
for large scale separation parameter $L_z/\ell_z$.

To derive Eq.~(\ref{TP21}) for the turbulent thermal energy $\langle E_\theta \rangle_{_{V}}$
in a convective  turbulence, we use budget equations~(\ref{C2})--(\ref{C3}) for $E_\theta$
and  $F_i$ in a steady-state for homogeneous turbulent convection:
\begin{eqnarray}
F_z &=& 2 C_{\rm F} \, t_{\rm T} \biggl[C_\theta \beta E_\theta
 - E_z \nabla_z \left(\overline{\Theta}^{({\rm m})} +\overline{\Theta}^{({\rm s})}\right)
 \nonumber\\
&& - {1 \over 2} F_r \left(\nabla_r  \meanU_{z}^{({\rm s})}\right)\biggr],
\label{TP15}
\end{eqnarray}

\begin{eqnarray}
F_r &=& - 2 C_{\rm F} \, t_{\rm T} E_r \nabla_r \overline{\Theta}^{({\rm s})} ,
\label{TP16}
\end{eqnarray}

\begin{eqnarray}
E_\theta &=&  - 2 C_{\rm p} \, t_{\rm T} \left[ ({\bm F} \cdot {\bm \nabla}) \overline{\Theta}^{({\rm s})} +
F_z\nabla_z \overline{\Theta}^{({\rm m})} \right] .
\label{TP17}
\end{eqnarray}
Using Eq.~(\ref{TP15}), we determine $\left\langle (F_z \nabla_z) \overline{\Theta}^{({\rm s})} \right\rangle_{_{V}}$.
Substituting the obtained expression into Eq.~(\ref{TP17}) averaged over the organized-structure volume,
we obtain expression for the turbulent thermal energy $\langle E_\theta \rangle_{_{V}}$ as
\begin{eqnarray}
&& \langle E_\theta \rangle_{_{V}} = 4 C_{\rm p} C_{\rm F}\, \ell^2 \biggl[A_z \, \left\langle\left(\nabla_z \overline{\Theta}^{({\rm s})}\right)^2\right\rangle_{_{V}}
\nonumber\\
&& \quad + A_r \left\langle\left(\nabla_r \overline{\Theta}^{({\rm s})}\right)^2\right\rangle_{_{V}} \biggr] ,
\label{TTP17}
\end{eqnarray}
where $A_r =E_r /E_{\rm K} = A_z = 1/3$ is the horizontal anisotropy parameter and
\begin{eqnarray}
\left\langle\left(\nabla_z \overline{\Theta}^{({\rm s})}\right)^2\right\rangle_{_{V}} &=&
A_\ast^2 \, \left\langle\left(\nabla_r \overline{\Theta}^{({\rm s})}\right)^2\right\rangle_{_{V}}
\nonumber\\
&=&{\pi ^2\over 2} \, \left({\Theta_0 \over L_z}\right)^{2} \, J_0^{2} (\lambda)  .
\label{TTP18}
\end{eqnarray}
To derive Eq.~(\ref{TTP18}), we use integrals given by Eq.~(\ref{BP1}).
Substituting Eq.~(\ref{TTP18}) into Eq.~(\ref{TTP17}), we obtain the turbulent thermal energy $\langle E_\theta \rangle_{_{V}}$
given by Eq.~(\ref{TP21}).

Averaging Eq.~(\ref{TP15}) for the vertical turbulent flux of potential temperature
over the volume of the semi-organized structure, we obtain that
\begin{eqnarray}
\langle F_z \rangle_{_{V}} =- \langle K_{\rm H} \rangle_{_{V}}  \, \nabla_z \overline{\Theta}^{({\rm m})}
+ \left\langle F_z^{({\rm W})} \right\rangle_{_{V}} + \left\langle F_z^{({\rm D})} \right\rangle_{_{V}} .
\nonumber\\
\label{TP18}
\end{eqnarray}
The first term on the RHS of Eq.~(\ref{TP18}) is the classical gradient transport term that describes the eddy heat diffusivity with
the turbulent diffusion coefficient $\langle K_{\rm H} \rangle_{_{V}}= 2 C_{\rm F}  \,\ell \, \sqrt{\langle E_{\rm K} \rangle_{_{V}}}$
that is given by
\begin{eqnarray}
\langle K_{\rm H} \rangle_{_{V}} = 2 \sqrt{3} C_{\rm F}  \, L_z \, U_{\rm D} \, \left({\ell_z \over L_z}\right)^{4 \over 3} \, \, {f_u^{1/2}(A_\ast)
\over  \left(1 - \langle \hat F \rangle_{_{V}}\right)^{1/3} } ,
\nonumber\\
\label{TP30}
\end{eqnarray}
where we use Eq.~(\ref{TP24}).
The second term on the RHS of Eq.~(\ref{TP18}) determines a secondary vertical turbulent flux,
$\langle F_z^{({\rm W})} \rangle_{_{V}} = -  C_{\rm F} \, t_{\rm T} \, \langle F_r \nabla_r \meanU_{z}^{({\rm s})} \rangle_{_{V}}$,
caused by an interaction of the horizontal turbulent flux $F_r$ of the potential temperature
with the local shear $\nabla_r \meanU_{z}^{({\rm s})}$ of the semi-organized structures,
where the horizontal turbulent flux $F_r$ of the potential temperature is given by Eq.~(\ref{TP16}).
Using Eqs.~(\ref{TP14}) and Eq.~(\ref{BP5}),
we obtain the ratio $\langle F_z^{({\rm W})} \rangle_{_{V}} /\langle F_z \rangle_{_{V}}$ as
\begin{eqnarray}
{\left\langle F_z^{({\rm W})} \right\rangle_{_{V}} \over \langle F_z \rangle_{_{V}} } = {\pi^2 \over {\rm Pr}_{_{T}}}\, {1 - \langle \hat F \rangle_{_{V}} \over A_\ast^2 \, f_{F_{T}}(A_\ast)}    .
\label{TP19}
\end{eqnarray}
The last term on the RHS of Eq.~(\ref{TP18}) describes the Deerdorf turbulent flux, $\langle F_z^{({\rm D})} \rangle_{_{V}} = 2  C_{\rm F} \, C_\theta \,  t_{\rm T} \, \beta \left\langle E_\theta \right\rangle_{_{V}}$, caused by potential temperature fluctuations.
Using Eqs.~(\ref{TP24}) and~(\ref{TP14}), we determine
the ratio $\langle F_z^{({\rm D})} \rangle_{_{V}}/\langle F_z \rangle_{_{V}}$ as
\begin{eqnarray}
&& {\left\langle F_z^{({\rm D})} \right\rangle_{_{V}} \over \langle F_z \rangle_{_{V}}} =  C_{\rm p} \, C_\theta \, C_{\rm F} \,
\left({\ell_z \over L_z}\right)^{2} \,  \left(1 - \langle \hat F \rangle_{_{V}}\right) \, f_D(A_\ast) ,
\nonumber\\
\label{TP20}
\end{eqnarray}
where function $f_D(A_\ast)$ is given by Eq.~(\ref{AP9}) in Appendix~\ref{Appendix:A}.
It follows from Eqs.~(\ref{TP19}) and~(\ref{TP20}) that $|\langle F_z^{({\rm W})} \rangle_{_{V}}|  / |\langle F_z \rangle_{_{V}}| \ll 1$ and $| \langle F_z^{({\rm D})} \rangle_{_{V}}| / |\langle F_z \rangle_{_{V}}| \ll 1$.
To derive Eq.~(\ref{TP19}), we use the following equation:
\begin{eqnarray}
&& \left\langle \left(\nabla_r \overline{\Theta}^{({\rm s})}\right) \,  \left(\nabla_r \meanU_z^{({\rm s})}\right) \right\rangle_{_{V}}
= {\lambda^2 \over 2 R^2} \,  \overline{\Theta}_{0} \, \overline{U}_{z0} \, J_0^2 (\lambda)
\nonumber\\
&& \quad= {\pi^2 \over L_z^2 \, A_\ast^2} \, \Theta_{\rm D} \, U_{\rm D} .
\label{BP5}
\end{eqnarray}
Equations~(\ref{TP18})--(\ref{TP20}) allow us to derive Eq.~(\ref{TP31}).

For the derivation of equations in this Appendix, we used the following integrals:
\begin{eqnarray}
&& \int_0^1 X \, J_0^2 (\lambda X) \,dX = \int_0^1 X \, J_1^2 (\lambda X) \,dX
\nonumber\\
&& =\int_0^1 X \, J_2^2 (\lambda X) \,dX=  {1 \over 2} J_0^2 (\lambda)={1 \over 2} J_2^2 (\lambda) ,
\label{BP1}
\end{eqnarray}

\begin{eqnarray}
\int_0^1 X \, J_0 (\lambda X) \, J_2 (\lambda X) \,dX=  {1 \over  \lambda^2} \left[1
- \left(1 + \half \lambda^2\right) J_0^2 (\lambda) \right]  .
\nonumber\\
\label{BP4}
\end{eqnarray}

\begin{eqnarray}
\int_0^1 J_0 (\lambda X) \, J_1 (\lambda X) \,dX=  {1 \over 2\lambda} \left[1 -J_0^2 (\lambda) \right] ,
\label{BP3}
\end{eqnarray}

\nocite{*}
\bibliography{paper-convection-PF-references}

\end{document}